\documentclass[12pt,twoside]{amsart} 
\usepackage[papersize={210mm,297mm},left=20mm,right=50mm,top=20mm,bottom=20mm]{geometry}
\usepackage{graphicx}
\usepackage{mathrsfs}
\usepackage{color}

\usepackage[colorlinks=false]{hyperref}
\hypersetup{linkcolor=slategray,urlcolor=slategray,citecolor=slategray}
\usepackage[preserveurlmacro]{breakurl}

\newcommand{\E}{{\rm e}}
\newcommand{\I}{{\rm i}}
\newcommand{\dd}{{\rm d}}

\newcommand{\pmni}{\par\medskip\noindent}
\newcommand{\pbni}{\par\bigskip\noindent}

\newcommand{\beq}{\begin{equation}}
\newcommand{\eeq}{\end{equation}}

\newcommand{\sfrac}[2]{{\textstyle \frac{#1}{#2}}}

\newcommand{\del}{\partial}

\newcommand{\cB}{{\mathcal B}}

\newcommand{\cG}{{\mathcal G}}
\newcommand{\cH}{{\mathcal H}}
\newcommand{\cS}{{\mathcal S}}
\newcommand{\cT}{{\mathcal T}}

\newcommand{\scD}{{\mathscr D}}
\newcommand{\scE}{{\mathscr E}}
\newcommand{\scF}{{\mathscr F}}
\newcommand{\scX}{{\mathscr X}}

\newcommand{\bM}{{\mathbb M}}
\newcommand{\bN}{{\mathbb N}}
\newcommand{\bR}{{\mathbb R}}
\newcommand{\bZ}{{\mathbb Z}}

\newcommand{\al}{\alpha}
\newcommand{\de}{\delta}
\newcommand{\et}{\eta}
\newcommand{\ga}{\gamma}
\newcommand{\si}{\sigma}
\newcommand{\ph}{\phi}

\newcommand{\ps}{\psi}
\newcommand{\om}{\omega}

\newcommand{\etq}{{\bar\eta}}
\newcommand{\phq}{{\bar\phi}}

\newcommand{\psq}{{\bar\psi}}

\newcommand{\veps}{\varepsilon}

\newcommand{\Ga}{\Gamma}
\newcommand{\Ps}{\Psi}
\newcommand{\Om}{\Omega}
\newcommand{\Tr}{{\rm tr}}

\newcommand{\escale}{\Omega}
\newcommand{\ez}{E}

\newcommand{\hopt}{{\tt t}}
\newcommand{\Sp}[1]{{\bf #1}}

\newcommand{\sLa}{\scX}

\newcommand{\ann}{a^{\phantom{*}}}
\newcommand{\cre}{a^{*}}

\newcommand{\MatsB}{\bM_B}
\newcommand{\MatsF}{\bM_F}

\newcommand{\chilbi}{h_>}
\newcommand{\bok}{{\bf k}}
\newcommand{\bop}{{\bf p}}

\newcommand{\reftitel}[1]{{\sl #1}}

\newcommand{\neu}[1]{#1}

\usepackage{wrapfig}

\begin{document}

\title[Renormalization in Condensed Matter]{Renormalization in Condensed Matter: Fermionic Systems -- from Mathematics to Materials}
\author{Manfred Salmhofer} 
\address{ 
Institut f\" ur Theoretische Physik, Universit\" at Heidelberg,
Philosophenweg 19, 69120 Heidelberg, Germany}

\email{salmhofer@uni-heidelberg.de}

\date{\today}

\begin{abstract}
\noindent
Renormalization plays an important role in the theoretically and mathematically careful analysis of models in condensed-matter physics. I review selected results about correlated-fermion systems, ranging from mathematical theorems to applications in models relevant for materials science, such as the prediction of equilibrium phases of systems with competing ordering tendencies, and quantum criticality.
\end{abstract}

\maketitle

\section{Introduction}
The renormalization of quantum field theories will always be linked to Wolfhart Zimmermann, who made seminal, famous contributions that remain relevant to this day, both in the application to particle physics and in mathematics. Although Zimmermann developed and applied renormalization techniques mainly focused on relativistic quantum field theory, the techniques he introduced apply in great generality, and they continue to inspire works in quantum field theory, as well as analysis and probability, e.g.\ recently in rough path theory and stochastic partial differential equations \cite{Hairer}. 

Renormalization also plays an important role in any careful treatment of models of condensed-matter physics, which can be cast in the form of models of nonrelativistic quantum field theory (QFT). These models are less symmetric than those of particle physics, and therefore renormalizing them is in general more complicated. Both ultraviolet and infrared renormalization is necessary in any given model, but the main focus has been on the infrared because most of the physically interesting effects are tied to it. In the condensed-matter setting, the models under consideration are defined independently of renormalization. The latter plays the role of a technique to solve certain small-denominator problems, which appear naturally, and which are analogous to secular terms in classical mechanics. They show up as divergences in a straightforward perturbative treatment. The counterterms introduced to renormalize the expansion turn out to be finite. They also have a physical interpretation. Thus, as in relativistic QFT, renormalization is not some `perverse recipe to get finite numbers', but it is a well-defined and natural procedure to treat a given problem mathematically correctly, by posing meaningful physical conditions. 

The approach to renormalization taken here is based on the renormalization group (RG) of Kadanoff,  Wilson, and Wegner, i.e.\ the successive integration over degrees of freedom, and the study of a flow of effective interactions derived from it. The connection between Wilsonian RG and field-theoretical renormalization \` a la Zimmermann has been elucidated, and brought in a form where a strict and rigorous discussion of renormalization conditions as in BPHZ renormalization is possible. The method is, however, not tied to perturbation theory -- in fact it is one of the most powerful non-perturbative methods available at present. On the mathematical side, it has been used by people working in {\em constructive field theory}, i.e.\ the mathematically rigorous construction of the correlation functions of models that are given by a Lagrangian or Hamiltonian. Famous results achieved by this method are the construction of 
infrared $\phi^4$ theory in four dimensions, and of the Gross-Neveu model, done by Gaw\c edzki and Kupiainen \cite{GK1,GK2} and, independently, by Feldman, Magnen, Rivasseau, and S\' en\' eor \cite{FMRS1,FMRS2}.
Ba\l aban proved stability of four-dimensional Yang-Mills theory \cite{BalYM} and analyzed the low-temperature behaviour of nonlinear sigma models in three dimensions \cite{Balsigma}.
Brydges and Kennedy showed how to derive tree and forest formulas from Wilsonian RG equations \cite{BK}. This was further developed by Abdesselam and Rivasseau \cite{AR}. On the one hand, this plays a role in drawing the connection to BPHZ renormalization, and on the other, it has become the basis of many current nonperturbative studies.

Going beyond the formal perturbation expansion involves a number of things: control of the combinatorial growth in expansion techniques, and tracking the flow of coupling functions. There are Landau-pole-like phenomena also in condensed matter. There, they are known to generate symmetry breaking. \neu{This has been shown for mean-field models of superconductivity \cite{SHML}. For the standard (non-mean-field) models of many-body theory introduced in the next section, a mathematical proof of continuous symmetry-breaking remains open: the infrared-bound method of Fr\" ohlich, Simon and Spencer \cite{FSS}, which has so far been the basis of all rigorous proofs of continuous symmetry breaking, is not applicable here because these models are not reflection-positive. (Only in very particular cases, e.g.\ hard-core bosons on a half-filled lattice, reflection positivity holds in the quantum many-body systems, due to a mapping to a spin system.) The program to show symmetry breaking for quantum many-body models using mathematically rigorous RG methods is, however, well-advanced in the case of Bose-Einstein condensation, see \cite{BFKT1,BFKT2,BFKT3,BFKT4} and \cite{Feldman}.} 

Last but not least, the technique developed here can be applied directly to the study of models relevant for experiments. While realistic models based on the details of stochiometry, crystalline structure, and orbital configurations, have been coming into reach only very recently, prototypical models have been studied using the RG method, and important phenomena have been understood this way, among them the effects of competing interactions for phase diagrams of layered materials, self-energy effects, and also transport properties. 

In this short review I will discuss equilibrium aspects of models of interacting fermions, which are often also called {\sl correlated-fermion models}. The focus will be on models with a Fermi surface that is really an extended object, i.e.\ it does not reduce to a collection of isolated points. 
The case of an extended Fermi surface is the harder case for renormalization, and it is the generic physical situation. I will quote theorems in an informal style, giving the reference to the full statements, which often require a much more detailed formulation.

\section{Condensed-Matter Models as Quantum Field Theories}

Quantum many-particle systems have a natural formulation in terms of second quantization. Although interesting mathematical work has also been done using Brownian motion techniques \cite{Ginibre,Koenig}, an important advantage of second quantization is that Bose and Fermi statistics, i.e.\ the bosonic symmetrization and fermionic antisymmetrization, are built into the formulation from the very start. To apply techniques of statistical mechanics, it has become standard in condensed-matter theory to rewrite the grand canonical traces over Fock space $\scF$ in terms of functional integrals. 

\subsection{Functional integrals for quantum many-body systems}

\neu{We first review the functional integral representation of the grand canonical ensemble of quantum many-particle systems, in an informal way, postponing a more careful mathematical discussion to the next subsection.} 
The partition function of a quantum many-body system governed by a Hamiltonian $H$ at inverse temperature $\beta = \frac{1}{k_B T}$ (where $k_B$ is Boltzmann's constant and $T \ge 0$ is the temperature) and chemical potential $\mu$ (which couples to the number operator $N$) is 
then rewritten as an integral, 
\beq\label{Zint}
Z
=
\Tr_{\scF} \; \E^{ - \beta (H - \mu N)}
=
\int 
\scD \bar\phi \, \scD \phi \;
\E^{- S(\phq,\ph)} \; ,
\eeq
over fields $\phi (\tau,x)$ and $\phq (\tau,x)$, which are complex variables in the case of bosonic particles, and Grassmann variables in the case of fermions. The fields depend on an additional variable $\tau$, the Euclidian time. (In our notation, the variable $x\in \sLa$ contains both the spatial coordinate and spin (or other internal) indices.
Given a Hamiltonian $H= H(\cre,\ann)$ in second-quantized representation, the exponent in the integral is the corresponding Euclidian action
\beq\label{formact}
S(\phq,\ph)
=
\int_0^\beta \dd \tau \;
(\bar\phi(\tau), \;  (\sfrac{\del}{\del \tau} + \mu ) \phi(\tau)
)_\sLa
- 
\int_0^\beta \dd \tau \;
H (\bar\phi(\tau) , \phi(\tau))
\eeq
with  
\beq
( \phi (\tau),\; \psi (\tau) )_\sLa
=
\int_\sLa \phi(\tau,x) \psi (\tau,x) \dd x \; ,
\eeq
where $\int_\sLa \dd x$ denotes summation over discrete indices and integration over continuous ones. 
The occurrence of a derivative with respect to $\tau$ in the action is a direct consequence of the canonical (anti)commutation relations of the bosonic (fermionic) creation and annihilation operators. 

A prototypical example is a gas of quantum particles interacting by a stable pair potential $v (x-y)$, 
\beq
\label{standHam}
H(\phq,\ph)
=
\frac12 (\phq, \; \scE \ph )_\sLa + \lambda (\phq\ph, \; v\, \phq\ph )_\sLa \; .
\eeq
Here $\lambda$ is a coupling constant. The kinetic operator $\scE$ in the quadratic part of the Hamiltonian can be $- \Delta$ or a suitable discretization thereof, or a hopping term for a lattice system, or come from a periodic Schr\"odinger operator. The interaction does not have to be quartic (corresponding to two-body interactions); three- and higher-body interactions can be included as well.

Bosonic fields are periodic, and fermionic fields are antiperiodic, under $\tau \to \tau+\beta$. 
Thus, at positive temperature $T$, the variable $k_0$ dual to $\tau$ is {\em discrete} because of the (anti-)periodic boundary conditions for the fields. 
In terms of the Matsubara frequencies $\omega_n = n \pi T$ for $n \in \bZ$,  
$k_0 \in \MatsB = \{ \omega_n : n$ even$\}$ for bosons, and $k_0 \in \MatsF =  \{ \omega_n : n$ odd$\}$ for fermions.

Finally, the integration measure in (\ref{Zint}) is
\beq\label{formeas}
\scD \bar\phi \, \scD \phi
=
\prod\limits_{\tau\in [0,\beta)\atop x \in \sLa}
\dd \bar\phi(\tau,x) \dd \phi(\tau,x)\;.
\eeq
Normalized expectation values of suitable linear operators $A$ on $\scF$ have the representation
\beq
\left\langle
A
\right\rangle 
=
\frac{1}{Z}\;
\Tr_{\scF} \; \left( 
A \; \E^{ - \beta (H - \mu N)}
\right)
=
\frac{1}{Z}\;
\int \scD \bar\phi \, \scD \phi \; 
\E^{- S(\phq,\ph)}\;
A(\phq,\ph) \; .
\eeq

\subsection{Mathematical treatment of functional integrals}
Both sides of (\ref{Zint}) require regulators in order to make sense mathematically. To make the trace on the left hand side well-defined, $\sLa$ must have a finite volume $|\sLa|$, and a short-distance regularization in $x$ (usually called ultraviolet regularization) is required, e.g.\ by making $\scX$ a lattice or restricting to a class of smooth functions. 
Introducing a finite volume is standard in statistical mechanics, since free energies and related quantities are extensive. The {\em thermodynamic limit} $|\sLa| \to \infty$ of expectation values and intensive quantities is expected to exist, and a nontrivial part of the proofs is to show that this is the case. It is the correct idealization to describe bulk systems and their phase transitions. It will also be taken below. 

The spatial ultraviolet (UV) regularization is often left in place (this will also be done here), since condensed-matter models often have a natural formulation on a lattice, be it by definition or due to restriction to a finite number of Bloch bands, which makes sense since crystals only exist up to a certain temperature, hence energy scale. Thus, for most applied questions of condensed-matter physics, the UV regulator can be kept. Removing it remains a fundamental issue, in order to treat true continuum systems. The renormalization problem entailed by it was solved in two spatial dimensions in \cite{BenfattoUV}, but remains open in three dimensions in the quantum field theoretical framework. 
The corresponding problem for the ground-state energy is the famous stability of matter problem, which was solved by Dyson and Lenard, and Lieb and Thirring (see \cite{LiebSeiringerbook}).

The right hand side of (\ref{Zint}) looks even more problematic, as it involves the formal product measure (\ref{formeas}), which does not really define a measure if $\tau$ is a continuous variable, even if $\sLa$ is just a finite set. This integral is really defined as a limit of a finite-dimensional integral corresponding to a discrete time variable $\tau$ with spacing, say, $\frac{1}{N}$. It is derived by using a Trotter formula for the exponential on the left hand side of (\ref{Zint}). 

It is nontrivial to prove that the time continuum limit $N \to \infty$ of the integral with a discretized version of the action 
(\ref{formact}) exists (and written in a form that can be used for further analysis).
For bosons, this was done recently in a series of papers \cite{BFKT1,BFKT2,BFKT3}, using a decimation-type renormalization group argument, as well as analysis of oscillatory integrals: the $\int \bar\phi \del_\tau \phi$ term has no positivity properties --- it is not even real.
For fermions, the situation is simpler -- a proof avoiding a multiscale analysis was given in \cite{PeSa}. The time continuum limit has also been shown to exist (in a form useful for the analysis of the thermodynamic limit) for a class of Bose-Fermi systems \cite{NiMS}. Typical Hamiltonians, such as (\ref{standHam}), are polynomials in the field operators, and the action $S$ is then a polynomial in the fields. We define the operator $Q$ by writing the quadratic part of the action $S$ in the form $(\phq, Q \phi)$, where the bilinear form now also includes time integrals: $(f,g) = \int \dd\tau (f(\tau),g(\tau))_\sLa$. For the choice (\ref{standHam}), 
\beq
Q(\tau,x,\tau',x') 
=
\delta(\tau,\tau') \; 
\left(
\delta (x,x') \; (\mu + \sfrac{\del}{\del \tau'})
-
\scE_{x,x'}  
\right) \; ,
\eeq
where the deltas denote the appropriate choice of Kronecker or Dirac delta functions, depending on the index, and we define the interaction $V$ to be the remaining part of the action $S$:
\beq
S= (\phq, Q \phi)
+ V(\phq,\ph) \; .
\eeq
We assume that $C=Q^{-1}$ exists. It is then useful to rearrange the integral for the partition function in terms of a normalized Gaussian measure
\beq
\dd\mu_C (\phq,\phi) 
=
Z_0^{-1}
\E^{-(\phq, Q \phi)_\sLa} \;
\scD \bar\phi \, \scD \phi
\eeq
where $Z_0$ is a normalization factor, formally given by 
$\det (\pi Q)$ for bosons, and $\det Q^{-1}$ for fermions.
$Z_0$ does not depend on the interaction, and it cancels out in normalized expectation values. To obtain expectation values of products of fields, it is convenient to introduce source terms, hence study the generating functional of the correlation functions (moments)
\beq
\begin{split}
Z (\etq,\eta)
&=
\hphantom{Z_0}
\int
\scD \bar\phi \, \scD \phi
\mkern15mu
\E^{-S(\phq,\phi)} 
\mkern19mu
\E^{(\etq, \phi)_\sLa - (\et,\phq)_\sLa} \;
\\
&=
Z_0 
\int \dd\mu_C (\phq,\phi) \;
\E^{-V (\phq,\phi)}\;
\E^{(\etq, \phi)_\sLa - (\et,\phq)_\sLa} \;
\end{split}
\eeq
In order to take the thermodynamic limit, one studies the connected correlation functions (cumulants), which are generated by 
\beq
W (\etq,\et)
=
- \log Z (\etq,\et) \; .
\eeq

\subsection{The Hubbard model}

In statistical mechanics, a prototypical model is the Ising model. Because it is so easy to define, but nevertheless exhibits phase transitions and other statistical mechanical phenomena in a simple, but nontrivial way, the Ising model is often called the {\em drosophila} of statistical mechanics. In a similar vein, I would like to call the Hubbard model the
{\em glossina morsitans}\footnote{Glossinidae, a.k.a.\ Tse Tse flies, were first described by 
the German scientist Christian Rudolph Wilhelm Wiedemann (1770-1840)} of correlated fermion models. 
The Hubbard model was introduced by several researchers in solid state physics and quantum chemistry in the 1950s, as a model of correlated fermions. The simplicity of its Hamiltonian stands in some contrast to the many opinions that have since been voiced about its possible equilibrium phases. Indeed, the reduction to a minimal set of ingredients in this model helps to bring out most clearly the major difficulties of many-fermion theory, and it is still only incompletely understood. I think this makes the above nickname all the more fitting.\footnote{Citation from Wikipedia: 
Most tsetse flies are physically very tough. 
Houseflies are easily killed with a fly-swatter but it takes a great deal of effort to crush a tsetse fly.}

Consider a weighted graph $\Lambda$, viewed as a collection of points $x$ and edge weights $\hopt_{x,x'}$ between points of the lattice, satisfying
$\hopt_{x',x} = \overline{\hopt_{x,x'}}$. In second-quantized formulation, the kinetic term of the Hubbard model  is the Hamiltonian 
\beq\label{Hubbhupf}
H_0 = \sum_{x,x' \in \Lambda} \sum_{\si = \pm}
\hopt_{x,x'} \cre_{x,\si} \ann_{x',\si} \; .
\eeq
Since the interpretation of each summand is that a fermion hops from site $x'$ to $x$, the $\hopt_{x,x'}$  are called hopping amplitudes. The index $\si = \pm$ is spin index for spin-$\frac12$ fermions. We focus here on spin-independent hopping amplitudes, but the spin-dependent case is of great interest as well, in particular in connection to topological phases. The hopping amplitudes may be complex, e.g.\ if the system is in an external magnetic field. 

The interaction of the original Hubbard model is quartic, as in (\ref{standHam}), here given by 
\beq
V
=
U \sum_{x \in \Lambda} \cre_{x,+}\ann_{x,+} \cre_{x,-} \ann_{x,-}
=
\sfrac{1}{2}\; U  
\sum_{x \in \Lambda} n_x^2 
- \sfrac12 \; U\, N
\eeq
where $n_x = \cre_{x,+}\ann_{x,+} + \cre_{x,-} \ann_{x,-}$ is the local fermion density operator, and $N = \sum_x n_x$. 
If interpreted as a substantially simplified version of a screened Coulomb repulsion between electrons in local orbital states, it is natural to assume that $U>0$, i.e. the energy increases by $U$ if two fermions occupy the same site. 

In condensed-matter applications, the vertex set $\Lambda$ of the graph is a regular lattice in $d$-dimensional space, where $d \le 3$, and the simplest variant of the model has nonvanishing hopping amplitudes only between nearest neighbours on that lattice. 

\begin{figure}[h]
\includegraphics[width=0.4\textwidth]{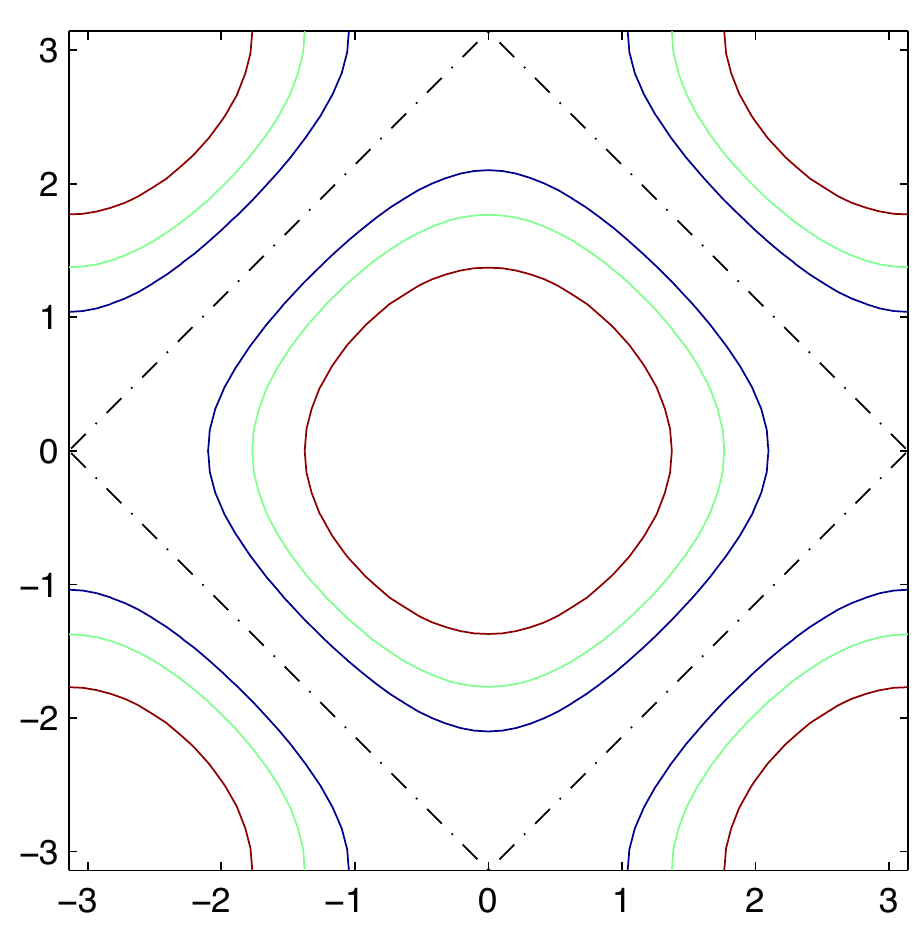}
\hfil
\includegraphics[width=0.4\textwidth]{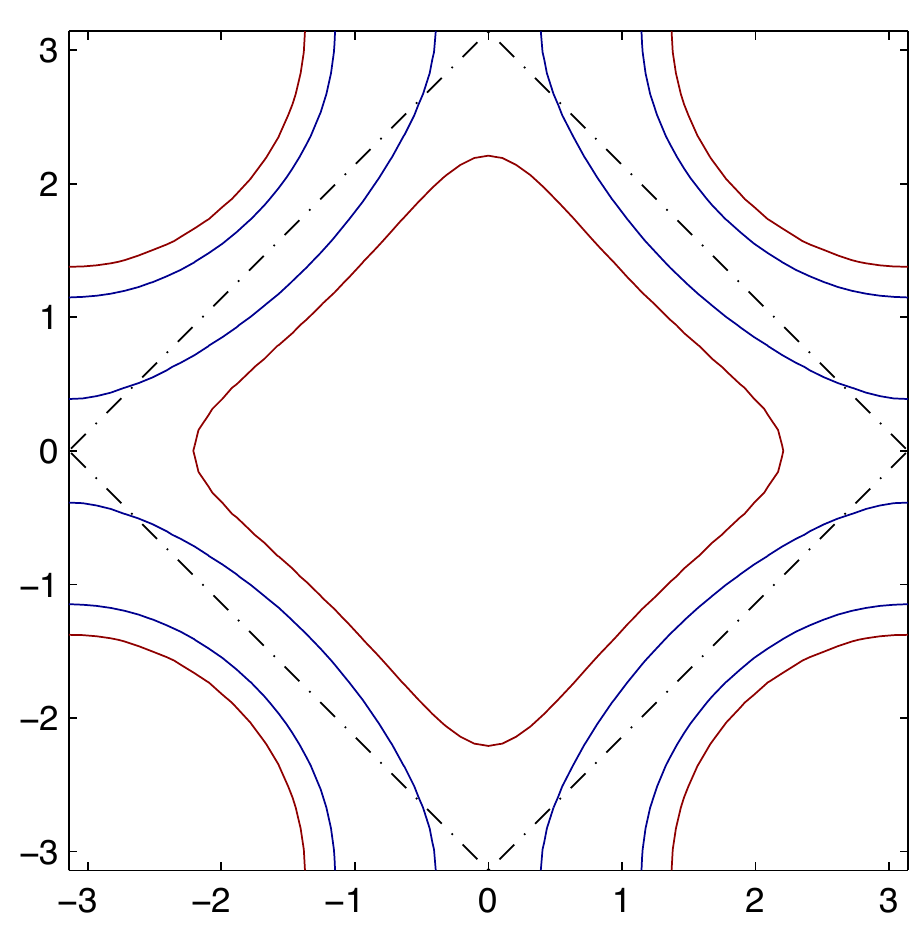}
\caption{The level lines (Fermi curves) of the two-dimensional band relation (\ref{ttpdisp}). Left: $\hopt' = 0$. Right: $\hopt'/\hopt < 0$.
.}
\label{fig1}
\end{figure}

It is obvious that the interaction can be generalized to include 
more than just an on-site repulsion, and this gives rise to additional effects, but here I restrict to the model on the two-dimensional square lattice with an on-site repulsion and consider hopping amplitudes $\hopt$ between nearest neighbours and $\hopt'$ between next-to-nearest neighbours. For an infinite lattice, momentum space is then given by the Brillouin zone $\cB =  \bR^2/ 2 \pi \bZ^2 \cong    [-\pi,\pi)^2$ with periodic boundary conditions, and the dispersion function for the single band given by $H_0$ is 
\beq\label{ttpdisp}
\epsilon(k) 
= 
- 2 \hopt (\cos k_1 + \cos k_2) 
- 4 \hopt' \cos k_1 \; \cos k_2 
\eeq
A collection of level sets of this function are shown for $\hopt' = 0$ and $\hopt' \ne 0$ in Figure \ref{fig1}. Depending on the density, every level line can be the Fermi surface of the model with $U=0$. For $\hopt'=0$, there is a particle-hole symmetry, which corresponds to the antisymmetry of $\epsilon (k) $ under $k \mapsto k + (\pi,\pi)$, and at half-filling ($\mu =0$), the Fermi curve is the square $D$ with corners $(\pm \pi, 0$ and $90,\pm \pi)$ (drawn dashed-dotted in the picture). This is an example of a perfectly nested Fermi surface, since its translate by $(\pi,\pi)$ is on top of $D$: $D+(\pi,\pi)=D$. Moreover, the saddle points of the dispersion function, $(\pi,0)$ and $(0,\pi)$, are on $D$.  For $\hopt' \ne 0$, particle-hole symmetry is absent, and there is no perfect nesting. There is approximate nesting in that the curvature of the Fermi surface vanishes on the Brillouin zone diagonal at a certain value of $\mu$. (This enhances incommensurate antiferromagnetic correlations.) Taking $\hopt' \ne 0$ disentangles the Van Hove filling,
\neu{where the zeroes of the gradient of $e$ lie on the Fermi surface,} from the filling at which \neu{the perfect} nesting occurs. 

If the hopping amplitudes are more general, but remain of finite range, $k \mapsto \epsilon(k)$ is analytic in $k$.
The theory developed in \cite{FST1,FST2,FST3,FST4}
applies to the Hubbard model at small $U$, with hopping amplitudes $\hopt_{x-x'}$ that depend only on $x-x'$ and have sufficient decay as $|x-x'| \to \infty $. They do not need to have finite range. In fact, the class of finite-range hopping Hamiltonians is too small to formulate and solve the renormalization problem mathematically. This means that, even if one starts with finite-range or even nearest-neighbour hopping in an interaction-free model, the interacting system will in general have a fermion dispersion that is not described by hopping amplitudes of any finite range. Similarly, it will become clear below that the effective low-energy-interaction, defined by the renormalization group flow, will not remain an on-site interaction. One can view phase transition points as those points where the effective interaction becomes long-range.  

\subsubsection*{Materials described by the Hubbard model}
The Hubbard model on the two-dimensional square lattice has received much attention since the late 1980s, because it was proposed as an effective one-band model for the {\em cuprates} materials with a layered structure that exhibit high-temperature superconductivity. They are called cuprates because the layers are composed of copper and oxygen atoms. The materials are three-dimensional, but the hopping amplitudes perpendicular to the planes is much smaller than that within the copper-oxide planes, so that modelling at not too low scales can focus on a two-dimensional system. The clean compounds are insulating antiferromagnets. The fermionic density can be changed by doping the material. At the density where the critical temperature $T_c$ for superconductivity is highest, a ``non-Fermi-liquid'' phase has been observed at temperatures above $T_c$  in the cuprates. \neu{The notion of a Fermi liquid will be explained in Section \ref{renliq}, and it will also be discussed there how one can define deviations from Fermi liquid behaviour in a precise way.}

While it is likely that the Hubbard model can serve as an idealization of the cuprates, in the sense that it qualitatively exhibits the basic phenomenology of these materials, as described very cursorily above, it certainly is not a realistic model for a quantitative \neu{description} of particular compounds: even the description of the hopping processes within the copper-oxide planes involves several orbitals, and gives rise to three- and more-band models. Moreover, the doping process also introduces impurities, which should be modelled by random terms in the Hamiltonian. One should therefore only look for the most robust features of the Hubbard model when bringing it into connection with cuprate physics. 

The Fermi surface of the cuprates, as determined from ARPES data, suggests that the Hubbard model may be a good description. The interaction of the fermions is expected to be relatively large---another challenge to all current theoretical efforts.  

The Hubbard model on a two-dimensional hexagonal lattice is the simplest model for single-layer graphene. The pure case (no dopants or other impurities) is the half-filled lattice, where the density is one fermion per lattice site. In this case, the zero level set of the fermionic dispersion function is not a curve, but a set of two points, and the dispersion function close to these points is, to a very good approximation, given by a Dirac operator. (The Dirac index comes about because the hexagonal lattice has two atoms per unit cell, so that the system is described by two bands). In this case, the electron density of states vanishes at the Dirac points, so that a small Hubbard interaction is irrelevant \cite{GM}.
The on-site interaction of realistic models for graphene is not small, however, and moreover, the fermions are not expected to screen the electromagnetic interaction well, so that a long-range interaction is more appropriate for the description. 

\section{Wilsonian renormalization group equations}
As mentioned, the generating functions defined above are formal, that is, they require a regularization in order to be mathematically well-defined. Specific regularizations for the class of models considered here, as well as their mathematical and physical motivation, will be discussed below. It turns out that flow equations obtained by introducing and varying a regulator that depends on a parameter $\escale$ (for instance, an energy scale) are also useful in exhibiting the combinatorial and analytical structure of the theory, and for doing bounds. 
In this section, we briefly review the general setup of the flow, to prepare for the applications that follow later. 

\subsection{Derivation of the RG equation} 
Let $\escale$ be a parameter on which $Q=Q_\escale$ depends differentiably, and such that $C_\escale = Q_\escale^{-1}$ exists. Then all generating functionals will also depend on $\escale$:
\beq
\E^{- W_{\escale}(\etq,\et)} 
=
\int \scD \bar\phi \, \scD \phi \; 
\E^{- (\phq,Q_{\escale}\phi) - V(\phq,\phi)}
\quad
\E^{(\etq,\phi) - (\et,\phq)}
\eeq
We assume, furthermore, that the interaction $V$ does not depend on $\escale$. Then
\beq\label{11}
\begin{split}
\escale \del_\escale\;
\E^{- W_{\escale}(\etq,\et)} 
&=
\int \scD \bar\phi \, \scD \phi \; 
\E^{- (\phq,Q_{\escale}\phi) - V(\phq,\phi)}
\;
(\phq, \dot Q_{\escale}\phi)\;
\E^{(\etq,\phi) - (\et,\phq)}
\\
&=
\int \scD \bar\phi \, \scD \phi \; 
\E^{- (\phq,Q_{\escale}\phi) - V(\phq,\phi)}
\;
\left(
\sfrac{\de}{\de \et}, \; \dot Q_\escale \sfrac{\de}{\de \etq}
\right)\;
\E^{(\etq,\phi) - (\et,\phq)}
\\
&=
\left(
\sfrac{\de}{\de \et}, \; \dot Q_\escale \sfrac{\de}{\de \etq}
\right)
\E^{- W_{\escale}(\etq,\et)} 
\end{split}
\eeq
Here $\dot Q_\escale = \escale \del_\escale Q_\escale$, 
the step from the first to the second line uses the standard differentiation rules for exponentials, and the step from the second to the third line uses that the differentiation with respect to the source terms can be taken out of the integral. 
Denoting the Laplacian in field space corresponding to some Operator $A$ by
\beq\label{Lap}
\Delta_{A} 
=
\left(
\frac{\de}{\de \et}, \; A \frac{\de}{\de \etq}
\right)
\eeq
this gives
\beq\label{Wfloeq}
\begin{split}
\dot W_\escale (\etq,\et)
&=
-
\E^{W_\escale (\etq,\et)} 
\Delta_{\dot Q_\escale} \; 
\E^{-W_\escale(\etq,\et)}
\\ 
&=
\Delta_{\dot Q_\escale} W_\escale(\etq,\et) 
-
\left(
\sfrac{\de W_\escale}{\de \et}, \; \dot Q_\escale \sfrac{\de W_\escale}{\de \etq}
\right) \; .
\end{split}
\eeq
Equations (\ref{11}) and (\ref{Wfloeq}) can be understood in simple terms by an analogy with heat flow: the integration (more precisely, the convolution) with a Gaussian measure generates the solution of a generalized heat equation ---  in the case of real fields and positive operators --- and the equation (\ref{11}) for the density $\E^{-W}$ therefore takes the form of a heat equation. The equation for the exponent $W$ then becomes a nonlinear heat equation. 

\subsection{Schemes}
A variety of schemes, that is, different forms of the renormalization group equation, have been used. 

\subsubsection{Polchinski's flow equation}
This is the variant of equation (\ref{Wfloeq}) obtained by setting $\etq = C^* \psq$, $\et=C\ps$. More precisely, the generating functional for the amputated connected Green functions,
\beq
\cG_\escale (\psq,\ps) 
=
 - (\psq, C_\escale^{-1} \ps) 
+ 
W_\escale(\psq \, C_\escale^{-1}, \; C_\escale^{-1} \ps) 
- \log Z_0(Q_\escale) , 
\eeq
satisfies
\beq
\dot \cG_\escale
=
- \Delta_{\dot C_\escale} \cG_\escale
+
\left(
\frac{\de \cG_\escale}{\de \et}, \; \dot C_\escale \frac{\de \cG_\escale}{\de \etq}
\right)
\eeq
This is the equation Polchinski first used to prove perturbative renormalizability of scalar field theory in a very elegant way \cite{Polch}.
This has since been developed into a versatile and powerful method of proving perturbative renormalizability of quantum field theories, by C.\ Kopper and coworkers (for a review and references see, e.g.\ \cite{Kopper}).

\subsubsection{The Wick-ordered flow equation}
\begin{figure}[h]
\includegraphics[width=0.8\textwidth]{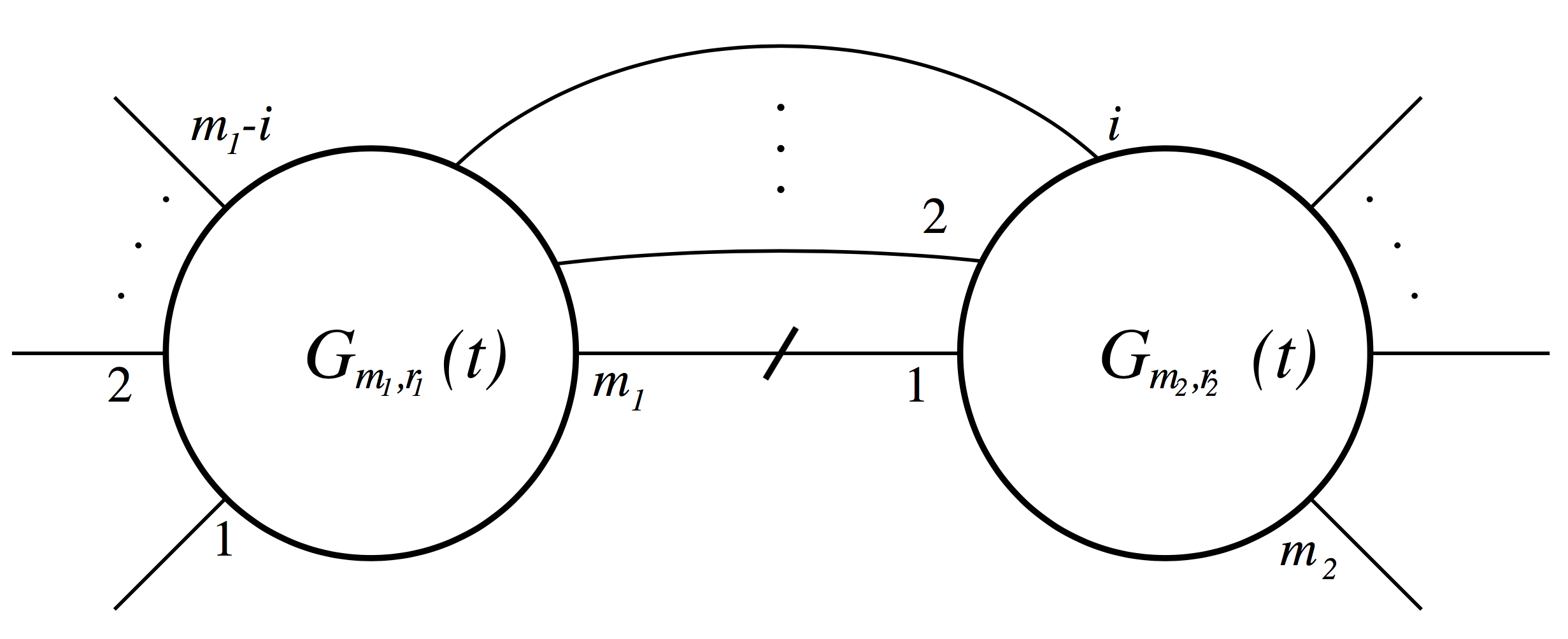}
\caption{The graph corresponding to the right hand side of the Wick ordered flow equation (\ref{Wickeq}). (Figure taken from \cite{msbook}.)}
\label{fig2}
\end{figure}

The generating functional for the Wick-ordered correlation functions is
\beq
\cH_\escale (\psq,\ps) 
=
\E^{\Delta_{D_\escale}} \cG_\escale
\eeq
where $D_\escale = C- C_\escale$ is the Wick ordering covariance. The resulting flow equation \cite{Sa98,msbook} is a quadratic differential equation, which can be written in the form \cite{MS-Hesselberg}
\beq\label{Wickeq}
\dot \cH_\escale (\Ps)
=
-
\left[
\E^{\Delta_{D_\escale}^{(1,2)}} \; \Delta_{\dot D_\escale}^{(1,2)} \; 
\cH_\escale (\Ps_1)\;  \cH_\escale (\Ps_2)
\right]_{\Ps_1 = \Ps_2 = \Ps} \; .
\eeq
The graphical depiction of the right hand side is shown in Figure \ref{fig2}. 

If $\escale$ plays the role of an infrared cutoff, the Wick ordering covariance $D_\escale$ is the covariance of the low-energy fields that have not been integrated over at flow scale $\escale$. Only quantities at scale $\escale$ or below enter  (\ref{Wickeq}). 

Wick ordering is particularly convenient for inductive proofs of renormalizability, since it simplifies the inductive scheme \cite{msbook}. Because the Wick ordering covariance is supported at scales below $\escale$, the Wick ordered equation also brings out the improvements to power counting that arise in systems with a non-nested Fermi surface in an especially transparent way \cite{Sa98}. 

\subsubsection{The flow equation for the irreducible vertex functions}
The generating functional for the one-particle irreducible (1PI) vertex functions is obtained in the standard way by a Legendre transformation
\beq
\Ga_\escale (\phq,\ph)
=
W_\escale (\bar H, H) - (\bar H,\phi) - (\phq,H)
\eeq
where $\bar H$ and $H$ are the functions of $(\phq,\phi)$ obtained by solving the equations
\beq\label{clafidef}
\phi= \frac{\de W_\escale}{\de \etq},
\qquad
\phq = \frac{\de W_\escale}{\de \et}
\eeq
for $\etq$ and $\et$. As in the case without the parameter $\escale$, it has to be checked whether this inversion is indeed possible. \neu{For bosonic fields, this requires strict convexity of $W_\escale$. For fermionic fields, it requires only the nondegeneracy of the quadratic part, due to the nilpotency of the Grassmann variables.} The functions $\bar H$ and $H$ also depend on $\escale$. By (\ref{clafidef}) and the standard relation 
\beq\label{Gaga2}
\frac{\de^2 \Ga}{\de \phq\de \phi} 
= \left(\frac{\de^2 W}{\de \etq \de \et}\right)^{-1}
\eeq
for the Hessians of $W$ and $\Ga$, the flow equation (\ref{Wfloeq}) implies the flow equation for the 1PI functions
\cite{Wetterich,SH}
\beq\label{1PIfloeq}
\dot \Ga_\escale 
=
(\phq, \dot Q_\escale \ph) 
-
\mbox{ Tr }
\left(
\dot Q_\escale 
\left(
\frac{\de^2 \Ga}{\de\phq\de\ph}
\right)^{-1}
\right)
\eeq

\subsection{Expansion in the fields}
\begin{figure}[h]
\includegraphics[width=0.7\textwidth]{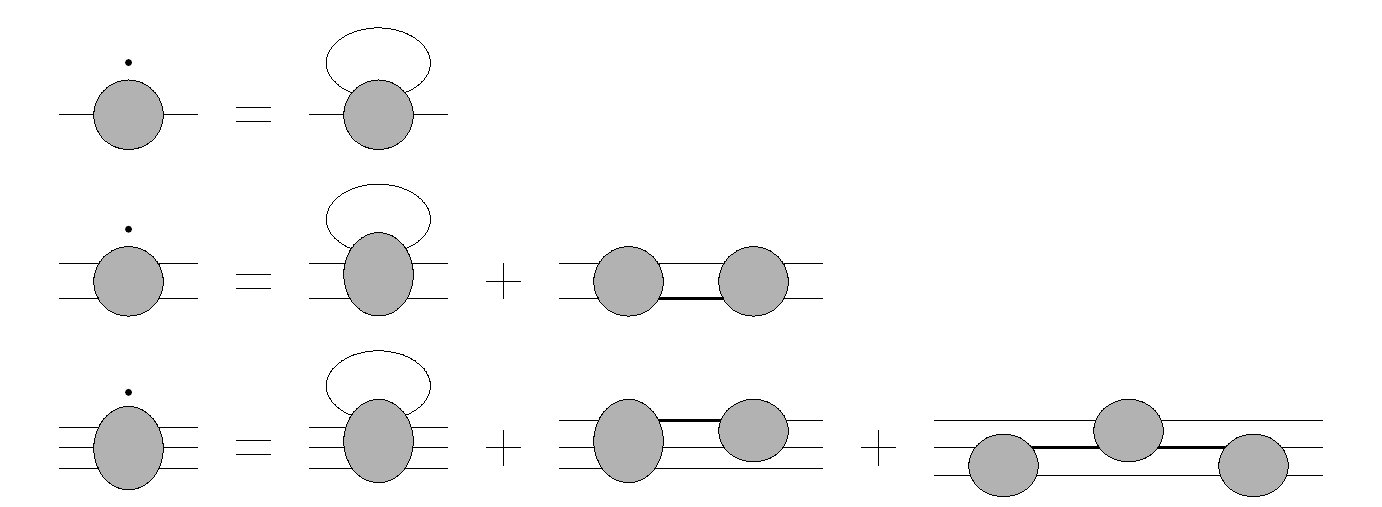}
\caption{Graphical representation of the first three equations in the RG hierarchy for the one-particle irreducible vertex functions.}
\label{fig3}
\end{figure}

The expansion of the various functionals in the fields gives the corresponding Green functions: $W$ generates the connected Green functions, $\cG$ the connected, amputated (by the free covariance) ones, $\cH$ the connected, amputated, and Wick ordered, ones, and $\Ga$ generates the one-particle irreducible (1PI) vertex functions. These expansions are all similar, and only the one for the 1PI functions is given here. For an even theory, where Green functions of an odd number of fields vanish, it reads 
\beq
\Ga_\escale (\Ps) = \sum_{m \ge 0} \Ga_\escale^{(m)} (\Ps)
\eeq
where $\Ga_\escale^{(m)} (\Ps)$ is of degree $2m$ in the fields $\Ps$. In a translation-invariant situation, where the fields  $\ps (K)= \ps_\alpha (k)$ can be labelled by $K=(k,\alpha)$, where $k$ combines momentum $\bf k$ and frequency $k_0$ and $\alpha$ is an internal index, e.g.\ spin, 
\beq
\Ga_\escale^{(m)} (\Ps)
=
\int \dd K_1... \dd K_{2m}\;
\de (\sum k_i) \;
\psq(K_1) ... \ps(K_{2m}) \; 
\ga_{m}^{\escale} (K_1, ... , K_{2m}) \; .
\eeq
The delta function enforces overall momentum and frequency conservation, and $\ga_m^{\escale}$ is the irreducible $m$-point function at scale $\escale$. Inserting this into (\ref{1PIfloeq}) and comparing coefficients in the fields results in an infinite hierarchy of equations. The first equation in the hierarchy is for $\ga_1^{\escale}$, which, by (\ref{Gaga2}), is the inverse of the full one-particle Green function (propagator) $G_\escale$. By (\ref{1PIfloeq}), it is of the form
\beq
\ga_1^{\escale}
=
\dot Q_\escale 
-
\Sigma_\escale
\eeq
where the scale-dependent self-energy $\Sigma_\escale$ is the solution of 
\beq\label{Sifloeq}
\dot \Sigma_\escale
=
\Tr \left(
S_\escale \; 
\ga_2^{\escale}
\right) \;,
\eeq
with an appropriate initial condition for $\Sigma_{\escale_0}$, where $\escale_0$ is the scale at which the flow is started (the choice of this scale depends on the regulator used). Here $S_\escale$ is the {\em single-scale propagator}
\beq\label{siscapr}
S_\escale 
=
- G_\escale \;
\dot Q_\escale\;
G_\escale \; ,
\eeq
and $\ga_2^{\escale}$ is the vertex function of the two-particle interaction. The full propagator then takes the form
\beq\label{fullpr}
G_\escale 
=
(Q_\escale - \Sigma_\escale)^{-1}
=
C_\escale \;
(1 - \Sigma_\escale C_\escale)^{-1}
\eeq
The right hand side of the flow equation for $\ga_2^{\escale}$ contains the six-point function $\ga_3^{\escale}$, etc. 
If every $\ga_m^{\escale}$ is represented by a vertex with $2m$ legs, these equations can be depicted as in Figure \ref{fig3}.

where the first equation represents (\ref{Sifloeq}), and each heavy internal line denotes a factor  $G_\escale$, the thin internal line a factor $S_\escale$. 

\subsection{Truncations} 
Setting $\ga_m^\escale = 0 $ for $m$ bigger than some $m_0$
allows to solve the equations recursively in terms of $\Sigma_\escale$ 
and $V_\escale$ (in principle). 
Already the truncation $\ga_3^\escale = 0 $ (or slight modifications thereof)
gives a nontrivial system of integro-differential equations
for the four--point vertex function$V_\escale (k_1,k_2,k_3)$
and the selfenergy  $\Sigma_\escale (k)$
\cite{SH,MSHMS}. Truncations of hierarchies are ubiquitous in many-body theory, e.g.\ the quantum Boltzmann equation of quantum kinetic theory is a truncation of the  quantum BBGKY hierarchy. The present truncation is on a higher level because it keeps the connected two-particle correlation function $\gamma_2^\Omega$, so the two-particle correlation does not factorize.

\subsection{RG equation for the many-fermion system}
For a many-fermion system of spin-$\frac12$ fermions, such as the Hubbard model, and under the assumption that $U(1) \times SU(2)$--invariance is unbroken, 
\beq
\ga_1^{\escale} (K,K') =\de_{\al,\al'} \; (\I k_0 - \ez(\Sp{k}) - \Sigma_\escale (k) \chi_\escale (k))
\eeq
(Here $\chi_\escale$ is the regulator function used to define the flow, specified in our applications below.)
Similarly, $U(1) \times SU(2)$--symmetry implies that 
$\ga_2^{\escale} (K_1, \ldots, K_4) $ is determined by the vertex $v_\escale (k_1,k_2,k_3)$
that describes the interaction of particles with spin conservation (for details and a derivation, see \cite{SH}). They satisfy the flow equation
\beq\label{eq:RGlevel2}
\begin{split}
\dot \Sigma_\escale( p )
&=
\frac12
\int \dd l\
S_\escale(l)\ 
\Bigl(
v_\escale( p, l, p) - 2 v_\escale(p, l, l)
\Bigr)
,\\
\dot v_\escale ( p_1, ..., p_4 ) 
&=
(\cT_{pp} + \cT_{ph, cr} + \cT_{ph, d})( p_1, ..., p_4 )
,
\end{split}
\eeq
where
\begin{align}
\cT_{pp} ( p_1, ..., p_4 )
&=
-\frac12 \int \dd l\ L_\escale( l, p_1+p_2-l )\ 
v_\escale(p_1, p_2, l)\  v_\escale(p_1+p_2-l, l, p_3),
\nonumber\\
\cT_{ph, cr}( p_1, ..., p_4 ) 
&=
-\frac12 \int \dd l\ L_\escale( l, p_1-p_3+l  )\ 
v_\escale(p_1, l, p_3)\  v_\escale(p_1-p_3+l, p_2, l),
\nonumber\\
\cT_{ph, d}( p_1, ... , p_4 ) 
&=
\frac12 \int \dd l\ L_\escale( l, p_2-p_3+l )\ 
\Bigl(
2 v_\escale(p_1, p_2-p_3+l, l)\  v_\escale(l, p_2, p_3)
\nonumber\\
&\qquad\qquad
- v_\escale(p_1, p_2-p_3+l, l)\  v_\escale(l, p_2, p_2-p_3+l)
\nonumber\\
&\qquad\qquad
- v_\escale(p_1, p_2-p_3+l, p_4)\  v_\escale(l, p_2, p_3)
\Bigr) \; .
\nonumber
\end{align}
Here
$L_\escale(p_1, p_2) = \frac{\partial}{\partial {\escale}}\ \big( G_\escale(p_1) G_\escale(p_2) \big)$.

\begin{figure}[h]
\begin{minipage}{0.45\textwidth}
\includegraphics[width=.9\textwidth]{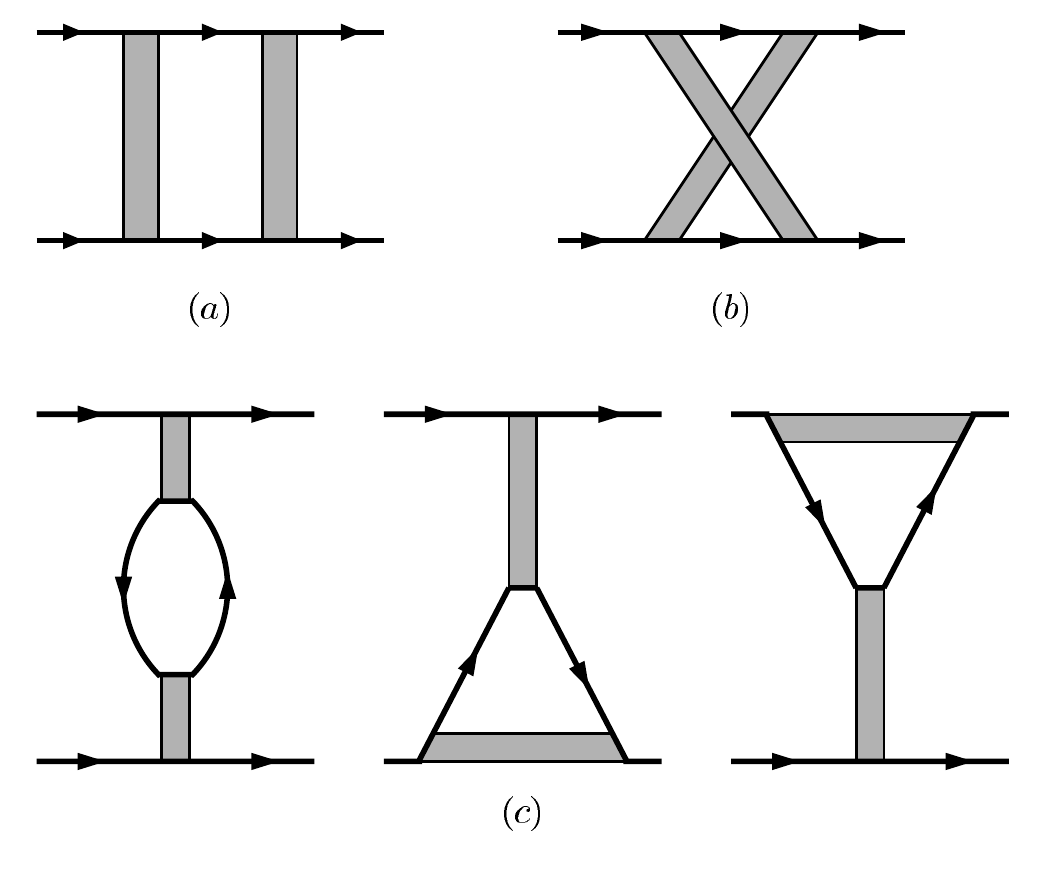}
\end{minipage}
\hfil
\begin{minipage}{0.38\textwidth}
\includegraphics[width=1.0\textwidth]{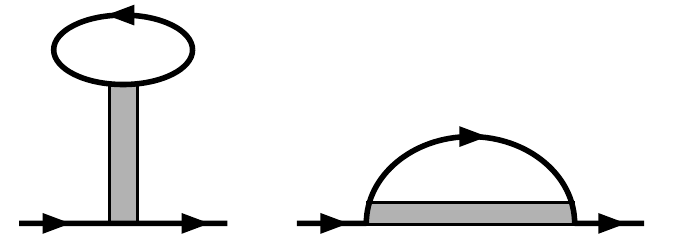}
\end{minipage}
\caption{The level-2 truncation of the 1PI hierarchy, for an $SU(2)$-spin and $U(1)$-charge invariant ansatz for the effective action. Spin is conserved along the heavy particle lines. Left: (a) the particle-particle contribution (b) the crossed particle-hole contribution (c) the particle-hole contributions. Right: the contributions to the fermionic self-energy. (Figure taken from \cite{SH}.)}
\label{fig4}
\end{figure}

Graphically, the terms on the right hand side of these equations can be represented by the diagrams shown in Figure \ref{fig4}.

Because of their structure, these equations are also called ``one-loop'' RG equations. It should be noted, however, that this is slightly deceptive --- the form of $L$ used here already involves particular two-loop contributions, which are essential for fulfilling Ward identities from the global symmetries, and which would not be captured in other truncations that also exhibit one loop in the graphical depiction. The equations can of course be solved iteratively in the interaction, thus generating a useful resummation of perturbation theory (corresponding to a simple truncation of the Brydges-Kennedy \cite{BK} formula). In lowest order, this really gives the bare one-loop diagrams of the theory. Because of the factor (-2) coming from the  fermionic minus sign and the spin sum in the RPA-type graph in (c), all diagrams in (c) cancel out for the bare Hubbard interaction. This is a special feature of the on-site interaction; it no longer holds if the interaction includes even next-to-nearest neighbour terms. 

\subsection{Relation to self-consistency equations}
An essential feature of the flow equations is that the vertex functions on the right hand side are not the bare ones. That is, one really has to solve an integrodifferential equation in order to obtain a solution. Restrictions obtained by retaining only one of the terms on the right hand side can be shown to converge to the solution of standard self-consistency equations \cite{SHML}. For instance, retaining only the particle-particle diagram (a) gives the result of the ``ladder resummation'' that indicates the instability of a normal Fermi liquid towards formation of Cooper pairs and a superconducting state, and retaining only the first diagram in (c) gives a standard random phase approximation (RPA). 

The full equations go significantly beyond Hartree-Fock, RPA, and quenched ladder approximations, and have become a workhorse in the analysis of correlated-fermion models:
it has been shown over the last two decades that if all terms are kept in this system of equations, it is possible to predict the possible phases of models for the cuprate and pnictide high-temperature superconductors, as well as a large number of other interesting materials. Some aspects of this will be discussed below; for a more detailed review, see \cite{MSHMS}.

\subsection{Flow parameters and regulators}

Above, $\escale$ was introduced simply as a parameter on which $Q=Q_\escale$, and hence the covariance of the free fields, $C_\escale = Q_\escale^{-1}$, depends. To be useful in analysis, the modification of $Q$ introduced by $\escale$ must have a regulating effect as well. The Wilsonian RG flow rearranges the functional integral such that degrees of freedom with correlations on short range are integrated first, and successively longer-range correlations are reached in the course of the flow. Operationally, this means that the covariance $C$, which in most applications has a slow decay, is modified by $\escale$ to one with a fast decay, whose decay length depends on $\escale$. We take the convention that the flow goes from an initial scale $\escale_0$ down to $\escale =0$, and the decay length increases as $\escale$ decreases.  This is the case if $\escale$ is chosen as an energy scale, or as an inverse length.  At the initial scale $\escale_0$, the covariance should either vanish, or at least the functional integral with covariance $C_{\escale_0}$ should be doable by a controlled approximation (e.g.\ a convergent expansion). 
As $\escale \to 0$, $C_\escale$ should converge to $C$, so that in the end, the full integral for the generating functionals has been performed. 

In the following, the setup of the flow and the most useful regulator functions are described, for the example of a many-fermion system with Fermi surface. 

A standard example of a regulator function $\chi_\Omega$ is a cutoff function in momentum space that vanishes in a vicinity of the Fermi surface. For $\escale_0 \ge \escale \ge 0$, we choose
$
Q_\escale (k) = Q(k) \; \chi_\escale (k)^{-1}
$
so that
\beq
C_\escale (k) = Q(k)^{-1} \; \chi_\escale (k) \; ,
\eeq
the full propagator (\ref{fullpr}) becomes 
\beq
G_\escale  (k)
=
\chi_\escale (k) \; 
(Q(k) - \chi_\escale (k) \Sigma_\escale (k))^{-1}
\eeq
and the single-scale propagator (\ref{siscapr}) becomes 
\beq
S_\escale (k)
=
-(Q(k) - \chi_\escale (k) \Sigma_\escale (k))^{-1}\;
\dot \chi_\escale (k) \; 
(Q(k) - \chi_\escale (k) \Sigma_\escale (k))^{-1}
\eeq
While having a well-defined $Q_\escale$ requires that $\chi_\escale (k) $ is nonvanishing (except at isolated points) because its inverse appears, the flow equations contain only the propagators, hence only $\chi_\escale (k) $, so one can take a limit to obtain a strict cutoff function that vanishes on open sets. 

The Fermi surface cutoff is obtained by taking 
\beq\label{FScutoff}
\chi_\escale (k)
=
\chilbi \left(
\sfrac{e(\bok)}{\escale}
\right)
\eeq
where $x \mapsto \chilbi(x)$ is a $C^\infty$ function that vanishes for $|x| \le 1$ and equals $1$ for $|x| \ge 2$. 

\begin{figure}[h]
\includegraphics[width=0.45\textwidth]{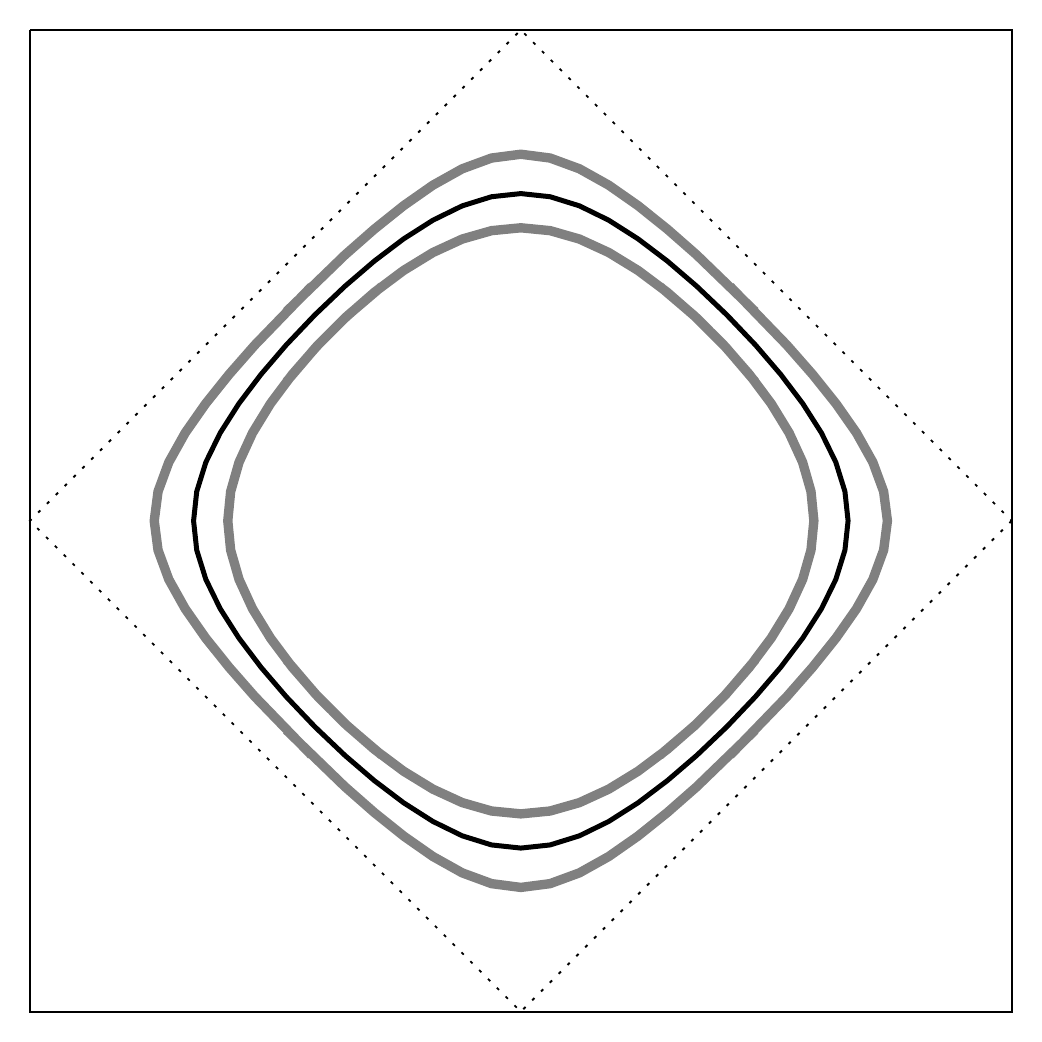}
\caption{The `momentum shell', i.e.\ the support of the scale derivative of the function (\ref{FScutoff}), in the Hubbard model with $\hopt' = 0$}
\label{fig5}
\end{figure}

Thus $\chi_\escale $ vanishes not only on the Fermi surface $\cS$, where $e(\bok) = \epsilon(\bok ) -\mu =0$, but in the open neighbourhood of $\cS$ where $|e(\bok)| < \escale$. Moreover, $\dot C_\escale$ and $S_\escale$ are supported on the thin shell in momentum space where $\chilbi'(e(\bok)/\escale)$ is nonzero. 
The shell is depicted in Figure \ref{fig5}, for the case of the Hubbard model with $\hopt' =0$ (the heavy line is the Fermi surface; the gray areas are the support of $\chilbi'(e(\bok)/\escale)$.

An even more stringent cutoff
\beq
\chi_\escale (k)
=
\chilbi \left(
\sfrac{e(\bok)^2+k_0^2}{\escale^2}
\right) \; ,
\eeq
which also restricts the Matsubara frequencies away from a small interval around zero, has been used frequently in mathematical studies. 

\begin{figure}[h]
\includegraphics[width=0.6\textwidth]{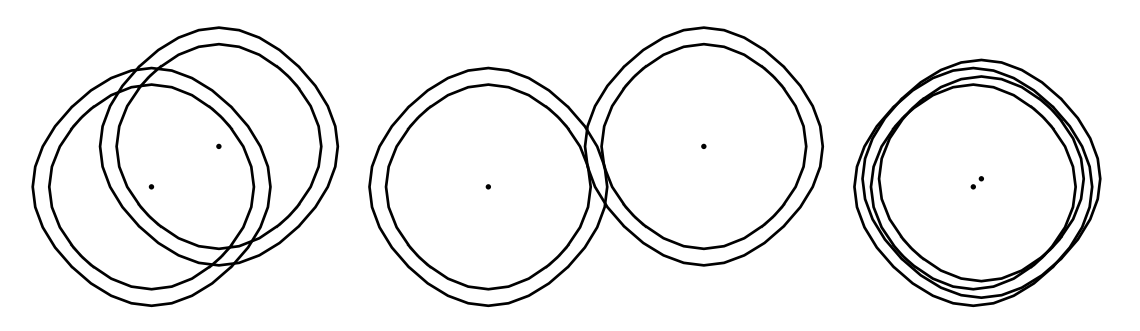}
\caption{The intersection of a thin momentum shell around the Fermi surface with its translates, for the Hubbard model with $\hopt' = 0$. The translating momentum $\bop$ connects the two centers, indicated by points. Left: for $\bop$ not close to zero and not close to $2 \cS$, the intersection is transversal, so it has much smaller volume in momentum space than the momentum shell itself. Middle: for $\bop$ in a neighbourhood of $2\cS$, the intersection is tangential, but the nonvanishing curvature still leads to a restriction. Right: for $\bop \approx 0$, the translates overlap and there is no volume improvement. (Figure taken from \cite{msbook}.)}
\label{fig6} 
\end{figure}

The Fermi surface cutoff is useful and has been used in many works. For instance, it provides direct insight into the role of geometry for possible singular behaviour of contributions to  the effective two-particle interaction:
singularities of the latter can occur only at those momenta $\bop$ where there is no transversal intersection of the Fermi surface and its translate by $\bop$ (see Figure \ref{fig6}). For a detailed review of these effects, see \cite{SaRMP}.

When analyzing intersections, it is very important to keep in mind that, for the lattice models considered here, momentum space is periodic, and that this significantly changes the analysis of intersections, so that there can be additional relevant terms in the flow, the so-called {\sl Umklapp} terms. Indeed, they are absent only at small fermion densities, and close to half-filling, they are essential for most properties of the flow in the two-dimensional Hubbard model \cite{halboth,HSFR,zanchi}.

A strict Fermi surface cutoff also has disadvantages; for instance, it suppresses small-momentum-transfer excitations artificially.
For this reason, alternative regulators have been introduced.
One of them is the temperature-flow RG \cite{HSPRL,HSTflo}, where the temperature $T$ is taken as the flow parameter. In this case, the general setup introduced above remains unchanged, but no function $\chi_\Omega$ needs to be introduced -- one simply takes a derivative with respect to the temperature after rescaling the fields such that the interaction term is independent of temperature. Another is the ``$\Omega$-scheme'' introduced in \cite{HuSa}, where 
\beq 
\chi_\escale(\om)  = \frac{k_0^2}{k_0^2 + \escale^2}
\eeq
This soft regulator, which depends only on the frequency variable $k_0$, is convenient for analyzing ferromagnetism and other small-momentum excitations. Moreover, because it does not depend on the spatial momentum $\bok$, flows with a dynamically adjusting Fermi surface can be defined straightforwardly. 

\section{Renormalization of the Fermi surface and Fermi liquid theory}\label{renliq}

Renormalization of perturbation expansions plays an important role in the mathematical theory of Fermi systems. Although it had been known for a long time to experts that renormalization is necessary, it was for a long time swept under the rug, mainly because it does not play a role in second order calculations and resummations based on them.  
A mathematically rigorous theory of renormalization began with the work of Feldman and Trubowitz \cite{FT1} and Benfatto and Gallavotti \cite{BG} at the beginning of the 1990s. A first analysis of the superconducting state was given soon after in \cite{FT2}. In these works, renormalization of the Fermi surface consisted in a renormalization of the chemical potential $\mu$ because the models were taken as rotationally symmetric. In a crystal, this is a useful approximation at small densities, i.e.\ when the chemical potential is just above a band minimum, because there the band function can be approximated by a constant times $\bok^2$. However, in general, rotational symmetry is broken to a discrete subgroup in a crystal, so it is important to understand the more general case of systems with nonspherical Fermi surfaces, where renormalization with acounterterm requires it to be a function of the momentum variable $\bok$. The dependence on $\bok$ cannot simply be parametrized by a finite set of parameters, because one needs a function of momentum to describe the deformation of a level set. 

The full perturbative renormalization of convex, curved Fermi surfaces was done in \cite{FST1,FST2,FST3,FST4}; it will be reviewed briefly below. This also allowed to remove some long-standing `paradoxes' \cite{KohnLuttinger} about the limit of zero temperature in many-body perturbation theory \cite{FKST}. These problems were due to an incorrect renormalization procedure that neglected the momentum dependence of the quadratic term in the fields, and they disappear once renormalization is done correctly \cite{FKST}.

Because the self-energy $\Sigma (k_0, \bok)$ contains all essential information about the properties of single-particle-like excitations (usually called quasiparticles) in interacting fermion systems, its analysis is also essential for understanding the notion of a Fermi liquid, which has been discussed extensively in the context of high-temperature superconductivity. Landau's concept of a Fermi liquid is often described in textbooks in terms of systems at temperature zero,
namely as a persistence of a discontinuity in the fermionic thermal occupation number at the Fermi surface, but it has been known since the work of Kohn and Luttinger that the standard many-fermion models exhibit symmetry breaking via a superconducting state at low temperatures. Consequently, the textbook definition does not really apply to any system where the band function $e$ is inversion-symmetric ($e(-\bok) = e(\bok)$). A definition of Fermi liquid (FL)behaviour at positive temperatures, i.e.\ above the critical temperatures for symmetry breaking, was introduced in \cite{Sa98}, and it was shown there that a system is an FL in the sense of this definition, if the fermionic self-energy is sufficiently regular as a function of $k_0$ and $\bok$. The results of \cite{FST1,FST2,FST3,FST4} and \cite{Sa98} imply this regularity to any finite order in perturbation theory for a large class of many-fermion systems. 

In this definition, it is essential that the system is not considered merely at a fixed temperature, but in an interval of (low) temperatures, and for the model to behave like an FL, it needs to exhibit a certain behaviour as a function of temperature: some functions must be uniformly bounded as a function of temperature (not all of them can be, since the region where FL behaviour holds is itself temperature-dependent). 

The classification of Fermi systems according to the properties of the fermionic self-energy also allows to specify deviations from FL behaviour in a more precise sense than just marking `non-FL-states' by the `absence' of FL behaviour. It turns out that in general, the self-energy is not a differentiable function of $k_0$ near to $k_0 = 0$. A dependence of $\Sigma (k_0,\bok) \sim |k_0|^{\alpha}$ with $\alpha < 1$ for $k_0 \to 0$ and $\bok$ on the Fermi surface \neu{has been found in several cases.}  
The most famous example for this is the one-dimensional fermion system, which exhibits Luttinger liquid (LL) behaviour: there, the exponent $\alpha$ is a function of the coupling constant $g$ that determines the interaction strength. For sufficiently small $g>0$ (the sign meaning a repulsive interaction), $\alpha $ is analytic in $g$ \cite{BGPS}. 
Singular self-energies have also been postulated and studied in two dimensions, in systems with long-range interactions, e.g.\ fermions coupled to gauge fields. 

In the following, I review renormalization theory for Fermi systems, both in the case with a regular and with a singular Fermi surface.

\subsection{Small denominators}

From the point of view of analysis, the renormalization problem can be viewed, at least in weakly interacting systems, as a small-denominator problem. If the quadratic part of the Hamiltonian $\scE$ has spectrum $\veps_\alpha$, then the covariance has an eigenfunction expansion in which 
$$
C(k_0,\alpha)
=
\frac{1}{\I k_0 - \veps_\alpha + \mu}
$$
appears. When $T$ gets small, the denominator can get small, depending on $\mu$. 

Because the fermionic Matsubara frequency $k_0$ is an odd multiple of $\pi T$, it never exactly vanishes at $T>0$. Therefore 
{\em positive temperature poses a natural infrared cutoff in fermion systems}. It is only in the limit $T \to 0$ that infrared divergences arise in Feynman graph expansions. 
In a translation-invariant situation, $\veps_\alpha = \epsilon(\bok)$ with momentum $\bok$. 
When in addition, the limit of an infinite system is taken, momentum $\bok$ becomes continuous and $C(k_0, \bok)$ has a singularity for 
\beq
k_0 = 0
\quad
{\rm and}
\quad
\bok \in \cS = \{ \bok \in \cB: e(\bok) = \epsilon (\bok) -\mu = 0\} \; .
\eeq
The nature of the singularity depends on $\mu$. Generically,  the $\mu$-level set of $\epsilon$ will be empty (if $\mu$ is in a band gap), or it will be a set of codimension $1$, that is, a surface for $d=3$ and a curve for $d=2$. For simplicity, and in accordance with standard convention, I will refer to them as ``Fermi surfaces'' in all dimensions $d \ge 2$.

This singularity causes infinities in every order $r \ge 3$ of the straightforward perturbation expansion. When the temperature $T$ is positive, these divergences get regularized, but since temperature is a physical parameter, it is important to ask what properties are uniform in temperature, or how things depend on temperature. 
The unrenormalized expansion yields spurious nonuniformities in temperature just as a naive perturbation expansion in mechanics can lead to secular terms.
In the many-body models, the correct renormalization allows to construct an expansion that is free of such spurious singularities. 

\subsection{Regular Fermi Surfaces}
Let the dispersion relation $\bok \mapsto e(\bok) = \epsilon (\bok) - \mu$ of the infinite system be a $C^{2}$ function. 
The Fermi surface $\cS$ is regular if $\nabla e \ne 0$ on $\cS$, so that $\cS$ is a $(d-1)$-dimensional submanifold of $\cB$. 

In this case, renormalization can be done with a momentum-dependent counterterm for the quadratic part of the action,
\beq
\int \dd^d k \phq(0,\bok) F(\bok) \ph(0,\bok) \; .
\eeq
Inspection of low-order perturbation theory shows that
$F(\bok)$ is a function of $\bok$ that is {\em not} just proportional to $\veps(\bok)$. 
Therefore, in a first step, sufficient regularity of $F$ needs to be proven. 

The physical meaning of $F$ is that it fixes the Fermi surface. 
In other words, the bare model is changed by a counterterm that is a function of $\bok$ {\em and} of the function $e=\epsilon - \mu$ and the interaction function, to fix the Fermi surface to $\cS$ in every order of the expansion. 
In \cite{FST1}, this order-by-order construction of the counterterm function is given in all dimensions $d \ge 2$, and it is proven there that for a class of systems with the properties

\begin{itemize}
\item
the Fermi surface $\cS$ satisfies a weak ``no-nesting'' condition: there can be only tangencies up to a fixed finite order in the intersection of the Fermi surface with its translates by nonzero momenta

\item
the interaction $v$ satisfies a summability condition (i.e.\ it is short-range, but not necessarily finite-range)

\end{itemize}
that to all orders, the counterterm function is finite and $C^{1}$ in $\bok$, at zero temperature. The results of \cite{FST1} also imply a similar behaviour for the self-energy as a function of $k_0$ and $\bok$, so there is no anomalous exponent $\alpha < 1$ that would imply a deviation from FL behaviour. The precise statement is given in Theorem 1.2 of \cite{FST1}. 

The counterterm function $F$ contains the information about the deformation of the Fermi surface under the influence of the interaction, but only in an implicit way: the statement is that if the original dispersion function is taken as
\beq\label{implicit}
E(\bok) = e(\bok) + F (e, v, \bok)
\eeq
where $F$ is determined order by order, then the expansion is finite to all orders. To renormalize the model, one needs to show that for given $E$ and $v$, one can find $e$ such that Equation (\ref{implicit}) holds. This is an implicit equation for a function, not just one variable. It is shown in \cite{FST1}, that $F$ is also Fr\' echet differentiable in $e$, and that the map $e \mapsto e+F$ is locally injective, so that uniqueness of the solution holds, if a solution exists. The implications and the meaning of the conditions imposed are discussed in detail in the introductory sections of \cite{FST1}. 

In \cite{FST2,FST3, FST4}, the existence of a solution $e$ of (\ref{implicit}) is proven for the class of many-fermion models where the condition on the Fermi surface is strengthened to require that the Fermi surface has positive curvature everywhere. This class includes, e.g.\ the Hubbard model away from half-filling. The bulk of the proof consists in showing that the second derivative of the dispersion relation and the curvature of the Fermi surface stay bounded in this case (Theorems 1.1 and 1.2 of \cite{FST3}), and that a solution of (\ref{implicit}) can be constructed by an iterative procedure (Theorem 1.1 of \cite{FST4}). The proofs apply in all dimensions $d \ge 2$, but there are subtle differences between the cases $d=2$ and $d \ge 3$. The two-dimensional case is the most singular one, and it is known from explicit calculations in special cases \cite{Fuji} that the properties proven for $F$ do not hold for the self-energy $\Sigma$ --- while $F$, which depends only on $\bok$, is $C^2$ in $\bok$, $\Sigma$, which also depends on $k_0$, is not $C^2$ in $k_0$.

The inversion theorem that guarantees that a solution $e$ of (\ref{implicit}) exists now also allows for a physical interpretation of the counterterm function. Inserting the solution of the implicit equation, the counterterm function gives the deformation of the Fermi surface under the influence of the interaction. It is independent of the projection operators used to implement renormalization, and of the renormalization scheme used. 

\pbni
Based on the perturbative analysis, and on a sectorization method for the Fermi surface \cite{FMRT}, two-dimensional fermionic models were constructed by a combination of renormalization group and convergent single-scale perturbation expansions, in several works. 
In a series of ten papers, Feldman, Kn\"orrer, and Trubowitz, constructed a two-dimensional Fermi liquid {\em at zero temperature} \cite{FKT}. The class of systems considered there is not symmetric under spatial reflection $\bok \to -\bok$. One can prove that this asymmetry removes the instability towards Cooper pairing. The regularity analysis of the Fermi surface for this case is included in \cite{FST1,FST2,FST3,FST4}. It is harder than in the symmetric case because the antipode (the point ${\bf a}(\bok)$ where the normal vector to $\cS$ is opposite to the normal at $\bok$) is no longer equal to $-\bok$, and because the treatment of certain third-order tangencies  is nontrivial (the rightmost case in the picture). 

\begin{figure}[h]
\includegraphics[width=0.6\textwidth]{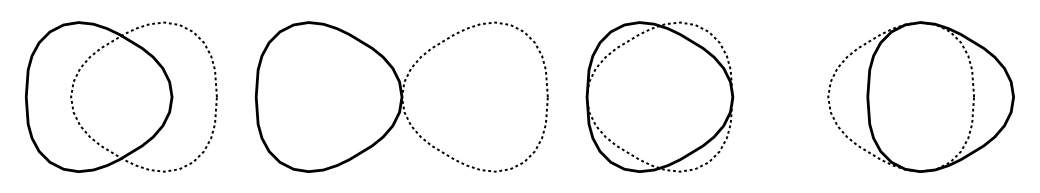}
\caption{Whether or not the Cooper interaction can become marginally relevant in the renormalization group flow is decided by the intersection of the Fermi surface $\cS$ with translates of its reflection, $-\cS$, because the constitutents of a Cooper pair have momenta $\bok$ and $-\bok$, both of which must be close to the Fermi surface to contribute to the flow at arbitrarily low scales. The figure shows the intersections for a band function $e$ for which $e(-\bok) \ne e(\bok)$, and different translation momenta $\bop$. From left to right: at generic momenta, the intersection is transversal; `at $2k_F$', the intersection is tangential but curvature suppresses the intersection volume; at $\bop=0$, the absence of reflection symmetry again leads to a suppression (in contrast to Figure \ref{fig6}); there are always translation momenta $\bop$ for which third-order tangencies must happen; these contributions remain suppressed if there are at most order $n$ tangencies for some fixed $n$. (Figure taken from \cite{FST2}.)}
\label{fig7}
\end{figure}

Another way to avoid the Cooper instability is to stay above the critical temperature for superconductivity, i.e.\ impose a condition $|\lambda| \log \beta < {\rm const.}$, where $\lambda$ is the strength of the interaction \cite{Sa98}.
The Fermi liquid condition of \cite{Sa98} was proven for a two-dimensional continuum system by Disertori and Rivasseau \cite{DR1,DR2}, and for the two-dimensional Hubbard model sufficiently far from half-filling by Pedra \cite{PeDiss} and Benfatto, Giuliani, and Mastropietro \cite{BGM}. For the three-dimensional case, only partial results exist.

\subsection{Singular Fermi Surfaces}

The results discussed in the previous section can be summarized qualitatively as follows: in two-dimensional Fermi systems with weak, short-range interactions, the self-energy is a regular function if the Fermi surface of the noninteracting Hamiltonian 

\begin{itemize}

\item
is weakly non-nested,\footnote{this is defined precisely in \cite{SFS1,SFS2}; essentially, at any point $p$ of the Fermi curve, the Fermi curve deviates from the tangent line at $p$ as a fixed minimal power of the distance from $p$} in particular it does not contain any segments that are straight lines

\item
is nonsingular, i.e.\ it does not contain any points $k$ with $\nabla \veps (k) = 0$

\end{itemize}

What happens when one of these conditions is violated? 
The case of long-range interactions is at the moment open mathematically. In the following, I will keep the condition that the interaction has short range, and present some results about {\em singular Fermi surfaces}, which contain points where $\nabla \veps (k) = 0$, so that they are no longer submanifolds of $\cB$. 

\subsubsection*{Van Hove Singularities} 

\begin{figure}[h]\includegraphics[width=0.3\textwidth]{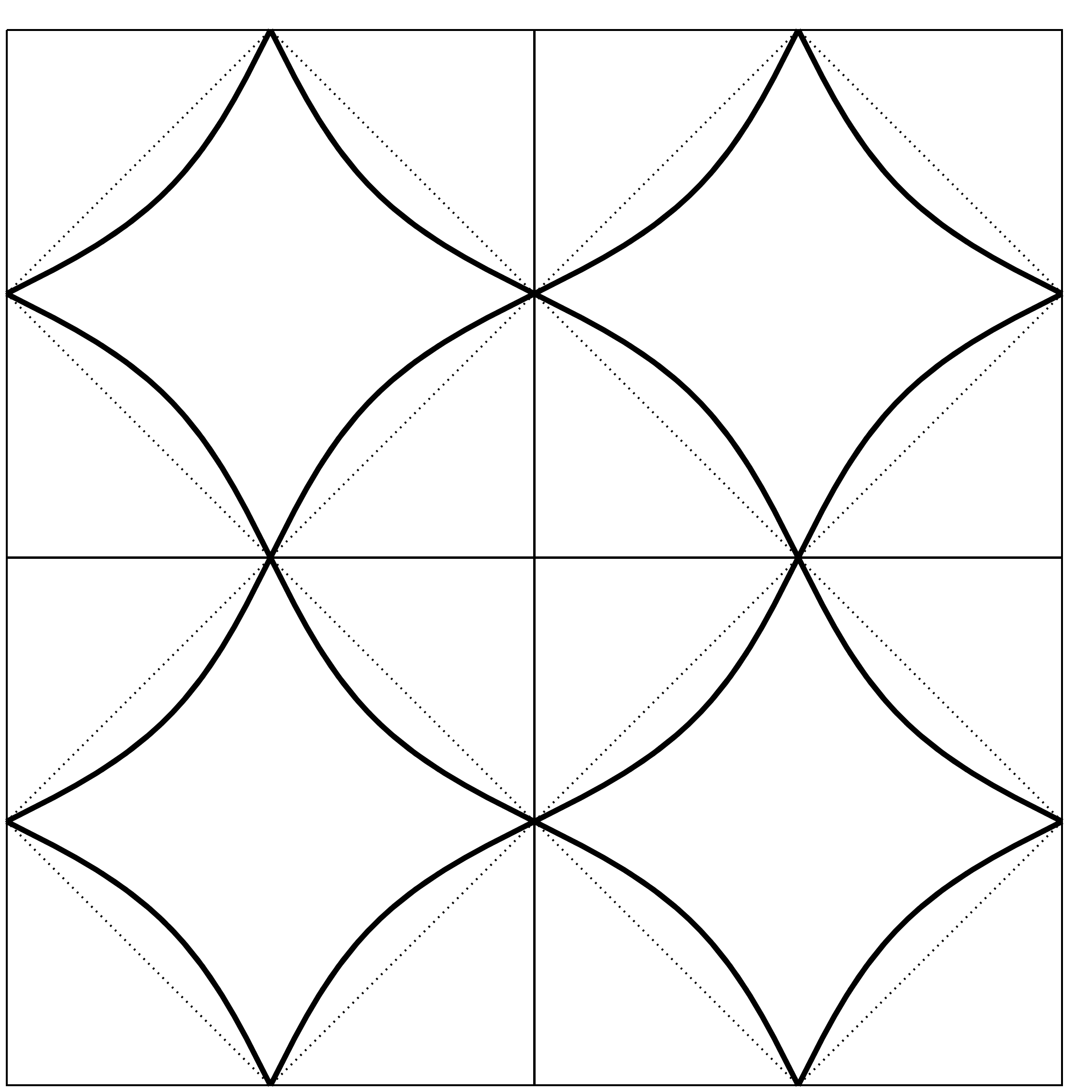}
\caption{The Fermi curve for (\ref{ttpdisp}) at van Hove filling.}
\label{fig8}
\end{figure}

Van Hove (1953) was the first to point out that generically, there must always be values of $\mu$ where $\cS$ contains zeroes of $\nabla e$. That is because $\bok \mapsto \epsilon(\bok)$ can be regarded as a Morse function. For a lattice system, momentum space is a torus, and it follows that the map $\bok \mapsto \epsilon (\bok) $ always has saddle points $\bok_s$, where $\nabla \epsilon (\bok_s) = 0$ and a rotation $R$ diagonalizes the Hessian at $\bok_s$, e.g.\ in two dimensions
\beq\label{epsilons}
\epsilon (\bok_s + R \bop) = E_{\rm VH} - \veps_1 p_1^2 + \veps_2 p_2^2 
+ O(|p|^3)
\eeq
with $\veps_i  \ge 0$. 
Generically,  $\veps_i > 0$, as the requirement that one of them vanishes introduces an additional condition. For the dispersion function (\ref{ttpdisp}) of the Hubbard model, one of these {\em Van Hove points} is at
$\bok_s = (\pi,0)$, with $E_{\rm VH} = 4 t'$ and 
\beq
\frac{\veps_1}{\veps_2} = \frac{1-2\theta}{1+2\theta}
\eeq
where $t'$ is chosen such that $0< \theta=-\sfrac{t'}{t}<1$. 
By symmetry, there is another Van Hove point at $(0,\pi)$. 
Geometrically, the Fermi curve has double points at the saddles of the dispersion, as shown in Figure \ref{fig8}, where translational copies of $\cB$ and $\cS$ are shown to exhibit the structure better.

\subsubsection*{Density-of-states effects}
The density of states
$
\rho (E) 
=
\int \sfrac{\dd^d k}{(2\pi)^d}\;
\delta (E - \epsilon (k))
$
has a {\em Van Hove singularity (VHS)} at $E = E_{\rm VH}$. This singularity is strongest in two dimensions, where
\beq
\rho(E) \sim \log\frac{W}{|E-E_{\rm VH}|} \; .
\eeq
(here $W$ is the bandwidth).
In higher dimensions, $\rho$ remains a bounded function, but divergences shows up in the derivatives of $\rho$.                         
The VHS have effects on the solutions of self-consistency equations. For instance, $\rho$ enters the BCS equation for superconductivity (SC)
\beq
\Delta
=
g \;
\Delta 
\int \dd E \; \frac{\rho(E)}{2\sqrt{(E-\mu)^2+\Delta^2}} \; 
\tanh \frac{\beta \sqrt{(E-\mu)^2+\Delta^2}}{2} \; .
\eeq
\pmni
In the absence of a VHS, that is, if $\rho$ is regular for $E \approx \mu$, the zero-temperature gap is proportional to $\E^{- \rho(\mu)/g}$. If there is a VHS, i.e.\  $\mu = E_{VH}$, then it becomes of order $\E^{-K/\sqrt{g}}$. VHS also enhance tendencies towards ferromagnetism (FM) in $d=2$. While ferromagnetism is absent in weakly coupled systems with regular Fermi surfaces, the presence of a VHS implies strong ferromagnetic correlations already at arbitrarily weak coupling, with a zero-temperature magnetization $\sim \E^{-\tilde K/J}$, where $J$ is the coupling strength of the ferromagnetic interaction. A comparison of these scaling behaviours suggests that if $J\sim U$ and $g \sim U^2$, then ferromagnetic and superconducting correlations can compete. This is indeed the case, as discussed below. However, simple simple gap equations or resummations are not sufficient to describe this competition.

\subsubsection*{The Van Hove scenario of high-Tc superconductivity}

Information about the Fermi surface of cuprate materials has been obtained from angle-resolved photoemission spectroscopy (ARPES). The results indicate that the Fermi surface is close to Van Hove points. They also hint at a suppression of the quasiparticle weight close to these points. In terms of the self-energy this would mean that the derivative of the self-energy gets large close to these points. It is a natural question whether this is caused by the nearby Van Hove singularities. Moreover, a striking result is that one of the two parameters $\veps_i$ of (\ref{epsilons}) is almost zero. Although such a nongenericity could be produced by adjusting hopping parameters on a particular lattice structure, this requires fine-tuning of the model parameters. In other words, its presence in a model of noninteracting fermions is not natural. There remains the possibility, however, that interaction effects lead to an extension of the Van Hove singularities in the sense that the gradient of $\epsilon$ is further suppressed by interaction effects. In an important early paper \cite{Dzy96}, Dzyaloshinskii studied the possibility of an extended Van Hove singularity with flat bands, marginal- and non-Fermi-liquid behaviour.
In \cite{FRS}, arguments for a pinning of the Fermi surface at the Van Hove points based on a simplified RG flow (the two-patch approximation, see below) were presented. The same approximation was used in \cite{IKK} to study the robustness of extended Van Hove singularities. 

\subsection{Rigorous RG study of the self-energy at Van Hove points}
The cuprate phenomenology, and the more general question in what cases FL theory fails even at weak coupling, motivate a more detailed study of the fermionic self-energy.
In \cite{SFS2}, we proved the following theorem. 

\pmni
{\bf Theorem. } {\sl 
Consider a two-dimensional Fermi system with a weak interaction. Assume that the Fermi curve of the system at zero interaction contains finitely many Van Hove points, and is non-nested (in a weak sense as above). Then

\begin{enumerate}

\item 
the renormalized perturbation expansion (where the Fermi surface is fixed by a counterterm) is finite to all orders, i.e.\ all correlation functions exist as formal power series in the coupling. 

\item
to every order in this expansion, the self-energy $\Sigma (k_0, \bok)$
is $C^1$ in $\bok$, uniformly in $k_0$ and $\bok$, 

\item
the $k_0$-derivative of $\Sigma$ is singular at $k_0 =0$
at the Van Hove point $\bok_s$ in the zero-temperature limit, and the quasiparticle weight $Z(\bok) = (1 + \I \del_{k_0} \Sigma (0,\bok))^{-1}$ vanishes at $\bok_s$. 

\end{enumerate}

}

\pmni

\begin{figure}[h]
\includegraphics[width=0.2\textwidth]{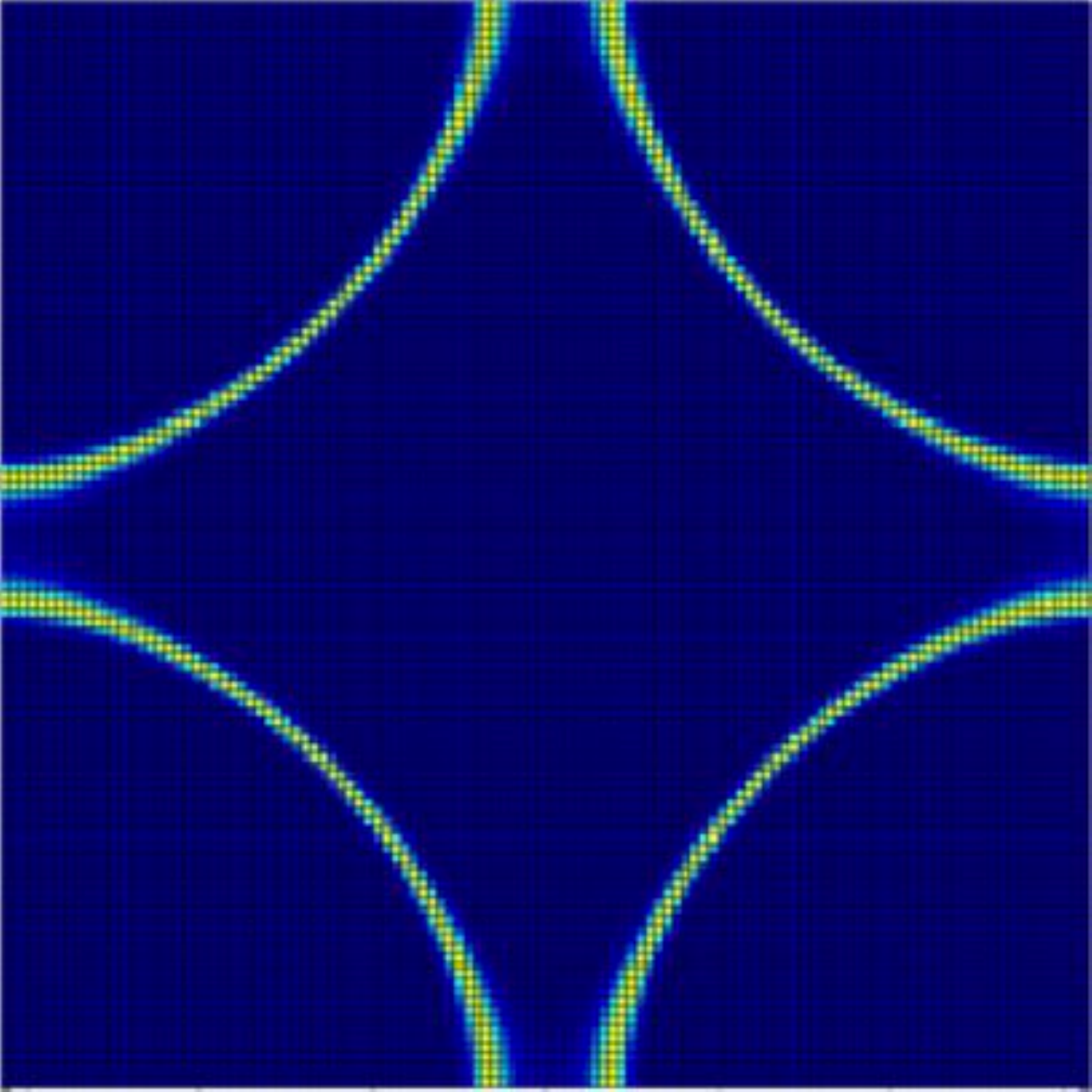}
\hfil
\includegraphics[width=0.2\textwidth]{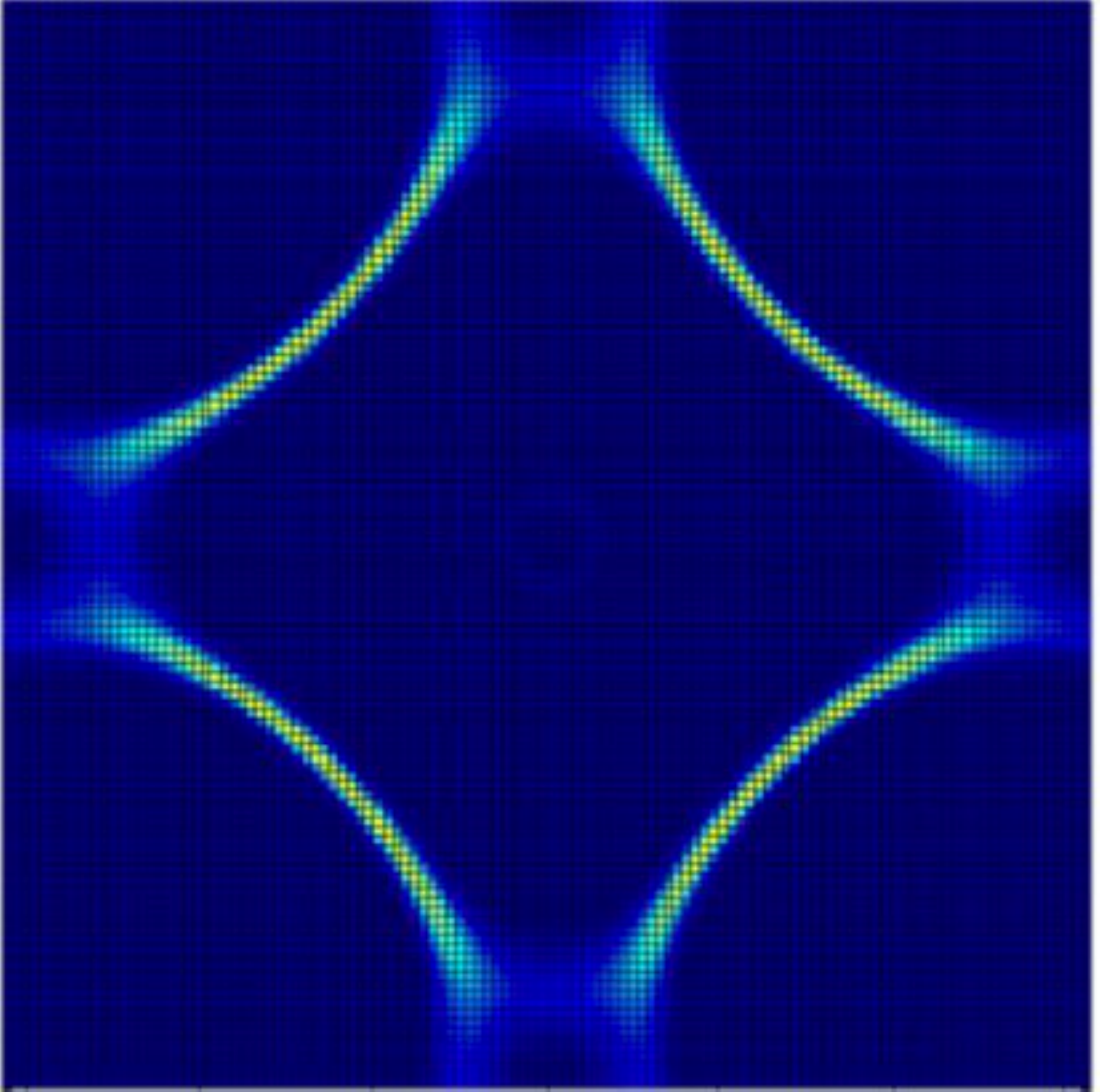}
\hfil
\includegraphics[width=0.2\textwidth]{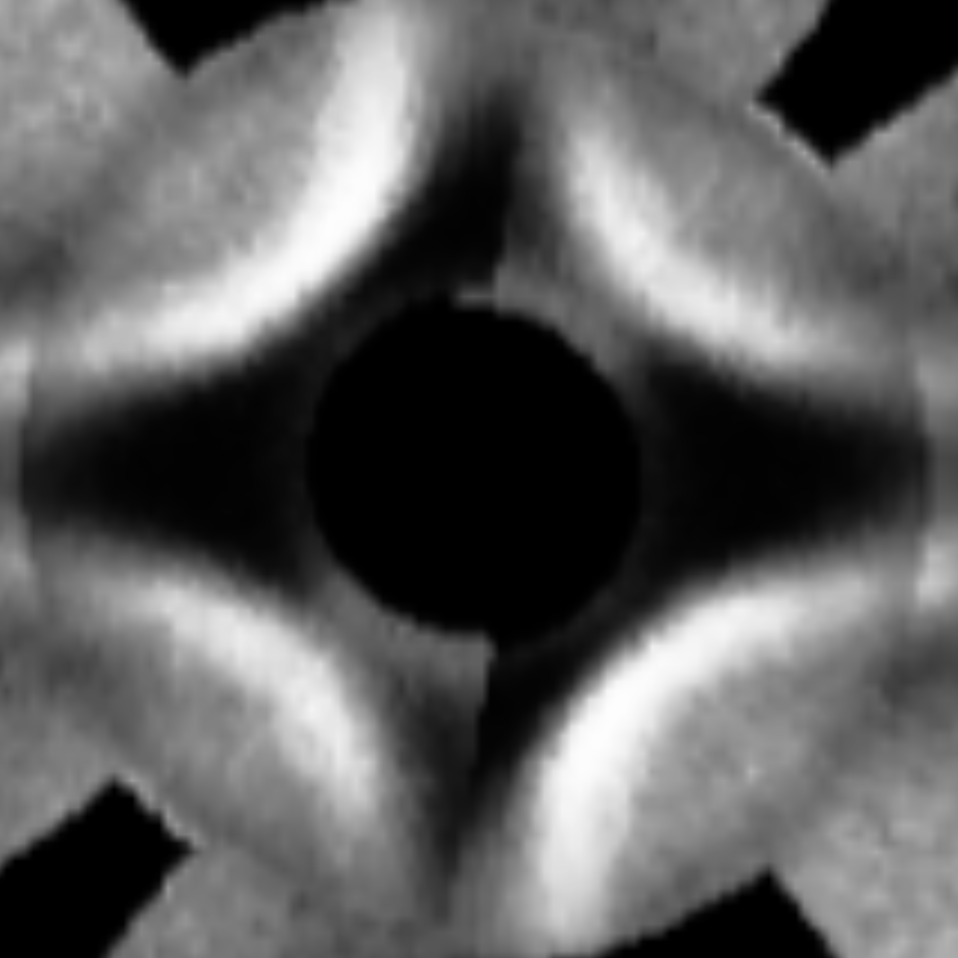}
\caption{Left: norm of the gradient of the fermionic occupation number, color-plotted over the first Brillouin zone of the square lattice, for the tight-binding band function (\ref{ttpdisp}). Middle: the similar plot, but including the singular behaviour of the self-energy in the interacting model. (Both images taken from \cite{SFS2}.) Right: ARPES curve of the underdoped Bi-2212 cuprate. (Image taken from Figure 1 in \cite{Bori02}, and cropped to show the same momentum region.)}
\label{fig9}
\end{figure}

Under mild assumptions on the optical matrix elements, ARPES measurements can be shown to probe the self-energy directly. Specifically, a symmetrization of the ARPES amplitude is proportional to the absolute value of the gradient of the occupation number density in $\bok$-space. The vanishing of $Z(\bok)$ at $\bok_s$ implies that this gradient vanishes at the Van Hove point, which suppresses the intensity of the ARPES signal, which then looks like a ``disappearance'' of the Fermi surface in the vicinity of the Van Hove points. Figure \ref{fig9} shows this effect of the $Z$ term in second order\cite{SFS2}; on the right hand side, experimental data from \cite{Bori02} are shown. For a detailed discussion of the relation of ARPES, cuprate band functions, and self-energies, see \cite{Bori06}.

The different regularity properties of the self-energy as a function of $k_0$ and $\bok$, as stated in the above theorem, can be understood by looking at the two-loop selfenergy
\beq
\Sigma_2 (q_0,\Sp{q})
=
\int_{k,p} 
\frac{1}{\I k_0 - e(\Sp{k})}\;
\frac{1}{\I p_0 - e(\Sp{p})}\;
\frac{1}{\I (q_0+p_0-k_0) - e(\Sp{q}+\Sp{p}-\Sp{k})} \; ,
\eeq
and studying its regularity properties by the same methods used to prove regularity in \cite{FST2}. The main point there is that the Fermi surface is not pointlike, but an extended object, and this leads to additional smoothing, as follows. One would like to avoid differentiating the denominators, because every derivative increases the power of the denominator, hence potentially creates singularities: the $L^2$ norm of the propagators diverges as $\log \beta$ for $\beta \to \infty$, and integrals of higher derivatives diverge like powers of $\beta$. The idea used in \cite{FST2} to prove regularity of the counterterm function to second order is to perform a change of variables to remove the $\Sp{q}$-dependence of the third propagator. The variables one can use to do this are the angular variables that parametrize the Fermi surface, which exist when the Fermi surface is not pointlike. Substituting $E=e(\Sp{q}+\Sp{p}-\Sp{k})$, all dependence on $\Sp{q}$ then
gets moved to the Jacobian of the change of variables. 
It turns out that this Jacobian has singularities, which are classified in \cite{FST2}, and it is shown there for regular, convex Fermi surfaces that the counterterm function is $C^2$. (It was noted there that the regularity holds only for the tangential derivatives, not for radial ones.) 

In the analysis of Fermi surfaces with Van Hove points, the Jacobian is singular when
$\Sp{q}$, $\Sp{p}$, and $\Sp{k}$ are Van Hove points, so that
 $\Sp{q}+\Sp{p}-\Sp{k}$ is also a Van Hove point. 
The non-nesting condition assures boundedness of the first derivative if $\Sp{q}+\Sp{p}-\Sp{k}$ is away from the Van Hove points. If $\Sp{p}$ and $\Sp{k}$ are such that {$\Sp{q}+\Sp{p}-\Sp{k}$} is close to a Van Hove point, then we proceed by taking the derivative of the propagator, to get
$$
\frac{\del}{\del \Sp{q}} \; \frac{1}{\I (q_0+p_0-k_0) - e(\Sp{q}+\Sp{p}-\Sp{k})}
=
\frac{\nabla e(\Sp{q}+\Sp{p}-\Sp{k}) }{(\I (q_0+p_0-k_0) - e(\Sp{q}+\Sp{p}-\Sp{k}))^2}
$$
and
$$
\frac{\del}{\del q_0} \; \frac{1}{\I (q_0+p_0-k_0) - e(\Sp{q}+\Sp{p}-\Sp{k})}
=
\frac{-\I }{(\I (q_0+p_0-k_0) - e(\Sp{q}+\Sp{p}-\Sp{k}))^2} \; .
$$
The numerator $\nabla e$ then provides a cancellation at the Van Hove point, and this makes it plausible that the function $\Sigma_2$ is more regular in $\bok$ than in $k_0$. In summary, the vanishing of $\nabla e$ at the Van Hove points 
cancels a singularity in the $\Sp{q}$-derivative, but not in the $q_0$-derivative, and one can show that there is no other cancellation in the $q_0$-derivative. 

The full proof of regularity \cite{FST3,FST4} and of the above Theorem \cite{SFS2} obviously requires more than just this argument, since the statement is not about second order contributions, but all orders of renormalized perturbation theory, and the combinations of loop momenta are much more general than just $\Sp{q}+\Sp{p}-\Sp{k}$. 
The proof avoids changes of variables and Jacobians in large graphs. Instead one uses the RG multiscale decomposition, the graph classification of \cite{FST1, FST3}, as well as length-of-overlap estimates, which generalize the overlapping loop estimates of \cite{FST1}, and in which the no-nesting hypothesis enters crucially, to obtain `volume improved' power counting bounds which are strong enough to accommodate for the loss in power counting entailed by one derivative with respect to $\Sp{q}$. 

This Theorem is proven for the situation where the Fermi surface is kept fixed by a counterterm. We have not proved an inversion theorem that generalizes the one for regular Fermi surfaces \cite{FST4} to the case of Fermi surfaces with Van Hove points. 

This regularity analysis also sheds light on a question one may have regarding the counterterm procedure. Could one pose a frequency-dependent counterterm to cancel this singularity? 
I know of no way of justifying counterterms for the frequency derivative, in contrast to the situation with momentum-dependent counterterms, hence no mathematically clear way of removing the singularity in the $k_0$-derivative. It is also not clear what this procedure would mean conceptually, since a model with a counterterm that depends more than linearly on $k_0$ does not correspond to a Hamiltonian of a many-body quantum system. 

The all-order statements  of \cite{SFS1,SFS2} might be proven non-perturbatively with the same technique and more careful bounds, but only nonuniformly in $\beta$
because, in general, the effective interaction increases under the RG flow, indicating the onset of symmetry-breaking. 
In the next section, I discuss a situation where in leading order, the effective interaction does not grow, due to a particular cancellation, and where the renormalized expansion may be controlled down to zero temperature.

\section{Competing order parameters and a quantum critical point}

The cuprate materials exhibit a number of different phases as a function of temperature and hole density. The antiferromagnetic and superconducting phases have been identified unambiguously, the nature of others remains to be clarified. The multitude of phases can be understood as a consequence of a competition of different ordering tendencies. Given the simplicity of the Hubbard model Hamiltonian, one may wonder how this model would explain all this, but RG analyses have shown that the phases arise in a natural way in the flow of scale-dependent effective interactions in this model. The advantage of the RG in this respect is that it does not require particular assumptions about ordering (as typically put in by an ansatz in mean-field studies), but that it allows for an unbiased analysis of the relative strength of competing interaction terms. The initial studies \cite{halboth,HSFR,zanchi} have been corroborated and extended in many further works; for a review, see \cite{MSHMS}.

\subsection{The composite-field expansion}

The above-mentioned studies were done using Fermi surface patching, i.e.\ a discretization of the dependence of all vertex and correlation functions as functions on the Fermi surface. 
An essential ingredient in a refined analysis of RG flows based on equation (\ref{eq:RGlevel2}) is a general representation for the effective interaction introduced in \cite{HuSa}:

$$v_\Om(p_1,p_2, p_3) = 2U + 
\left( v_{pp}^\Om + v_{ph,cr}^\Om + v_{ph, d}^\Om \right)(p_1,p_2, p_3)$$
with $\dot v_{pp}^\Om = \cT_{pp}$, $\dot v_{ph, cr}^\Om = \cT_{ph, cr}$,
$\dot v_{ph, d}^\Om = \cT_{ph, d}$.

These functions correspond to the following interaction types:

\begin{itemize}

\item
$\quad v_{SC}^\Om = v_{pp}^\Om$ to Cooper pair interactions

\item
$\quad v_{M}^\Om = v_{ph, cr}^\Om$ to spin interactions

\item
$\quad v_{K}^\Om = 2 v_{ph, d}^\Om - T_{34}v_{ph, cr}^\Om$ to charge density interactions

{\small (here $T_{34} v(p_1,p_2,p_3) = v(p_1, p_2, p_1+p_2-p_3)$ is a permutation operator that exchanges variables 3 and 4.)}

\end{itemize}

The idea in \cite{HuSa} is that the singularity in each of these functions develops as a function of a particular linear combination of momenta, namely those where the Fermi surfaces overlap maximally, so that a singularity can occur. \neu{Consequently,} these functions have the expansions
\begin{align}\label{eq:exchangeparametrisation}
v_{SC}^\Om ( p_1,p_2,p_3 )
&=
-
\sum_{m, n} f_m( \frac{ \bop_1 + \bop_2 }{2} - \bop_1 )\ D_{mn}^\Om( p_1 + p_2 )\ f_n( \frac{ \bop_1 + \bop_2 }{2} - \bop_3 )
\nonumber\\
v_{M}^\Om ( p_1,p_2,p_3 )
&=
\phantom{+}
\sum_{m, n} f_m( \bop_1 -  \frac{ \bop_1 - \bop_3 }{2} )\ M_{mn}^\Om( p_1 - p_3 )\ f_n( \bop_2 +  \frac{ \bop_1 - \bop_3 }{2} )
\nonumber\\
v_{K}^\Om ( p_1,p_2,p_3 )
&=
-
\sum_{m, n} f_m( \bop_1 +  \frac{ \bop_2 - \bop_3 }{2} )\ K_{mn}^\Om( p_2 - p_3 )\ f_n( \bop_2 -  \frac{ \bop_2 - \bop_3 }{2} )
\nonumber
\end{align}
Here $(f_m)_{m \in \bN}$ is an orthonormal system on momentum space. 
The singularity builds up in the  functions $D_{mn}$, $M_{mn}$ and $K_{mn}$, which couple the fermionic bilinears in the same way that an exchange boson couples to the corresponding fermionic bilinears, by a Yukawa-type term. As long as the vertex function of the boson-fermion coupling does not develop a singularity itself, it can in principle always be represented by an expansion in an orthonormal system. We chose an orthonormal system supported on momentum space rather than on the Fermi surface, because expansions in `Fermi surface harmonics' are tailored to the Cooper interaction term, and do not simplify the analysis of the other (particle-hole-type) terms. 

\begin{figure}[h]
\includegraphics[width=0.7\textwidth]{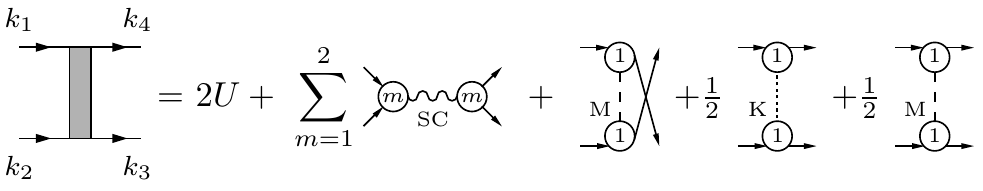}
\caption{Graphical representation of the form (\ref{eq:channel-splitting2}) of the scale-dependent effective fermionic interaction for the Hubbard model. Solid lines correspond to fermion fields; the dashed and wavy lines correspond to the interaction functions $D_{11}$, $D_{22}$, $M_{11}$ and $K_{11}$. As the graphs suggest, these correspond to the exchange of bosons, which can be made explicit by a Hubbard-Stratonovich transformation in the functional integral.}
\label{fig10}
\end{figure}

It is instructive to write the decomposition explicitly in terms of the fermionic bilinears, as follows. 
$V =  V_B +  V_K +  V_M +  V_C$, where $V_B$ is the bare Hubbard interaction, and
\begin{equation} \label{eq:channel-splitting2}
\begin{split}
V_K(\overline\psi,\psi) 
&= 
-\sfrac14 \int\sfrac{\dd^3{\ell}}{(2\pi)^3} \sum_{m,n=1}^\infty K_{m,n}(\ell)\, S^{(0)}_m(\ell) S^{(0)}_n(-\ell),
\\
V_M(\overline\psi,\psi) 
&= 
-\sfrac14 \int\sfrac{\dd^3{\ell}}{(2\pi)^3} \sum_{m,n=1}^\infty M_{m,n}(\ell)\sum_{j=1}^3 S^{(j)}_n(\ell) S^{(j)}_m(-\ell),
\\
V_C(\overline\psi,\psi) 
&= 
+\int\sfrac{\dd^3{\ell}}{(2\pi)^3} \sum_{m,n=1}^\infty D_{m,n}(\ell)\sum_{j=0}^3 \bar C^{(j)}_m(\ell) C^{(j)}_n(\ell)\; .
\end{split}
\end{equation}
The fermionic bilinears are 
\begin{equation}\label{eq:bilinears-def}
\begin{split}
S^{(j)}_m(\ell) &= \int\sfrac{\dd^3{q}}{(2\pi)^3} \;  f_m(\Sp{q}) \;\overline\psi_q^T \sigma^{(j)} \psi_{q+\ell}\;,\\
\bar C^{(j)}_m(\ell) &= \frac{\I}{2} \int\sfrac{\dd^3{q}}{(2\pi)^3} \; f_m(\Sp{q}) \; \overline\psi_q^T \sigma^{(j)} \overline\psi_{\ell-q}\;,\\
C^{(j)}_m(\ell) &= \frac{\I}{2} \int\sfrac{\dd^3{q}}{(2\pi)^3}\; f_m(\Sp{q})  \;\psi_q^T \sigma^{(j)} \psi_{\ell-q}\;,
\end{split}
\end{equation}
where $\psi(p) = (\psi_+(p), \psi_-(p))^T$, similarly for $\overline \psi(p)$, $\sigma^{(j)}$ are the Pauli matrices ($\sigma^{(0)} = 1$) and \(f_{m}\) are scale independent form factors, in particular 
\begin{equation}\label{formfactors}
\begin{split}
f_1(\Sp{q}) &= 1 \;,\\
f_2(\Sp{q}) &= \cos(q_x)-\cos(q_y)\;.
\end{split}
\end{equation} 
In the case $m=1$ we drop the subscript from the bilinear, writing $S^{(0)}$ for $S^{(0)}_1$.
This decomposition is represented graphically in Figure \ref{fig10}. 

We have kept the initial interaction separate in our decomposition so that the exchange propagators are zero at the beginning of the flow.

The interpretation of this ansatz for the effective interaction is that composite fields given by fermionic bilinears, namely density, Cooper pair, and spin fields, are coupled by functions 
$D_{mn}$, $M_{mn}$ and $K_{mn}$. The effective action also contains the kinetic term of the low-energy fermionic degrees of freedom, which includes the self-energy from the integration down to scale $\Omega$. The main idea of this decoupling is that singularities in $v$ that develop during the RG flow show up in the functions $D_{m,n}$, $M_{m,n}$ and $K_{m,n}$, which are formally similar to boson exchange propagators, while the form factors of the fermionic bilinears are regular functions, hence square-integrable, so that the expansion in $m$ and $n$ applies. Strictly speaking, boson propagators need to have positivity properties to preserve stability, but as long as the fermionic RG is used, this is, in fact, not necessary. Indeed, the results obtained for the functions $D_{mn}$, $M_{mn}$ and $K_{mn}$,  provide a test whether the effective action can really be bosonized. That singularities arise only as functions of the combinations $p_1 - p_3$, $p_2 - p_3$, and $p_1 + p_2$ can be strictly proven for small coupling functions,\cite{SaRMP,Sa98}, but in later stages of the flow it is an assumption. When entering the symmetry-broken phase, where some of $D_{mn}$, $M_{mn}$ and $K_{mn}$,  develop singularities, the form factors themselves may become singular and this channel decomposition is no longer accurate, and must be refined \cite{Eberleinfein}.
We shall, however, use (\ref{eq:channel-splitting2}) only in the symmetric phase, that is, we will stop the flow before any singularities arise; there, our approximations are justified, and our parametrization has the advantage of allowing to switch to a description in terms of bosonic order parameter fields by a straightforward Hubbard-Stratonovich transformation (provided the above-mentioned positivity holds). 

In general, all functions may depend both on the spatial momenta and on the Matsubara frequencies. We have made the approximation that the form factors are independent of the frequencies (see (\ref{formfactors})). In Ref.\ \cite{HuSa}, we verified that this approximation, and keeping \(K_{1,1}\), \(M_{1,1}\), \(D_{1,1}\) and \(D_{2,2}\), suffices for an accurate representation of the flow in the two-dimensional square-lattice Hubbard model. Thus we drop all other pairs $(m,n)$ from the sums in (\ref{eq:channel-splitting2}). All these exchange propagators except $K_{1,1}$ remain positive during the flow. The static part of \(K_{1,1}\) is positive, but it develops a pronounced negative minimum at nonzero frequency. 
The full density-density interaction also receives a contribution from $M_{1,1}$ because a local interaction $-\frac14 M(0)\int \Sp{S}(\ell)\Sp{S}(-\ell) \dd{\ell}$ can always be decomposed into a magnetic interaction $-\frac14 M(0)\int S^{(3)}(\ell)S^{(3)}(-\ell) \dd{\ell}$ and a density-density interaction $-\frac12 M(0)\int S^{(0)}(\ell)S^{(0)}(-\ell) \dd{\ell}$. 

We have also shown that keeping the frequency-dependence of the functions $K$, $M$ and $D$ requires also keeping the frequency-dependence of the fermionic self-energy, and taking into account their interplay in the RG flow \cite{MS-HuGiSa}.

\subsection{The case for `deconfined' quantum criticality in the two-dimen\-sional Hubbard model}

In many examples, quantum critical points (QCP) are shielded by some ordered phase, in the sense that quantum critical scaling does not continue all the way to zero scale, but it gets cut off due to some symmetry breaking at a very small scale. It is therefore an interesting question whether transitions exist where no such shielding occurs. The notion of {\sl deconfined quantum criticality} was introduced in \cite{Senthil}. The example chosen there is the transition from a N\' eel state to a valence-bond state in a two-dimensional spin-$\frac12$ antiferromagnet with the Hamiltonian $H = J \sum \Sp{S}_x \cdot \Sp{S}_{x'}$, where the sum runs over nearest-neighbour pairs and $J>0$. It is argued that a {\sl fractionalization} of the order parameters occurs, so that the transition is via a quantum critical point that has no simple description by a Ginzburg-Landau-Wilson (GLW) model, and that a local gauge symmetry emerges at the transition. This is made explicit by writing the spin at each point as $S_x^{(i)} = z_x^\dagger \sigma^{(i)} z_x$ with a complex spinor $z_x$. The decomposition introduces a redundancy because the $S_x^{(i)}$ are invariant under the local $U(1)$ transformation $z_x \mapsto \E^{\I \theta_x} z_x$, $\theta_x \in \bR$. At the transition, the $z$ fields can deconfine due to Skyrmion excitations of the N\' eel order parameter field, the order parameter fractionalizes and a gauge theory emerges. 

In the following, I discuss the evidence for a QCP in the Hubbard model at Van Hove filling. Within the approximations applied in our studies it appears to be an example of a `deconfined' transition, in that it remains unshielded. The interesting point in view of the above discussion is that it comes directly from a singular fermionic self-energy, and its dynamics is not driven by topological excitations, but by the vanishing of the quasiparticle weight at the Van Hove points. This also suppresses magnetic and superconducting correlations, and hence gives a picture that differs from the one from \cite{Senthil} sketched above:  the magnetic and superconducting order parameter fields are the composite fields given in (\ref{eq:bilinears-def}). They can of course always be studied, but the suppression of the order parameters at the QCP simply means that instead of getting fractionalized, these composite fields never really develop any important size and correlations at the transition point. (Moreover, although the composite magnetic fields are invariant under the standard local $U(1)$ number symmetry transformation of the fermions, this is not a symmetry of the model because the kinetic term of the fermion fields is not invariant.) 

First evidence for the QCP in the Hubbard model was found in \cite{HSPRL}, 
as a transition from $d$-wave superconductivity to ferromagnetism, when introducing the {\em temperature-flow RG} \cite{HSTflo}. Because momentum-space cutoffs artificially suppress ferromagnetism, this 
had not been found in earlier studies. 
In \cite{HSPRL}, we used the standard level-2 RG equation (\ref{eq:RGlevel2}) with a Fermi surface sectorization (patching), the approximation by frequency-independent vertices, 
and we neglected the self-energy.

\begin{figure}[h]
\includegraphics[height=0.3\textwidth]{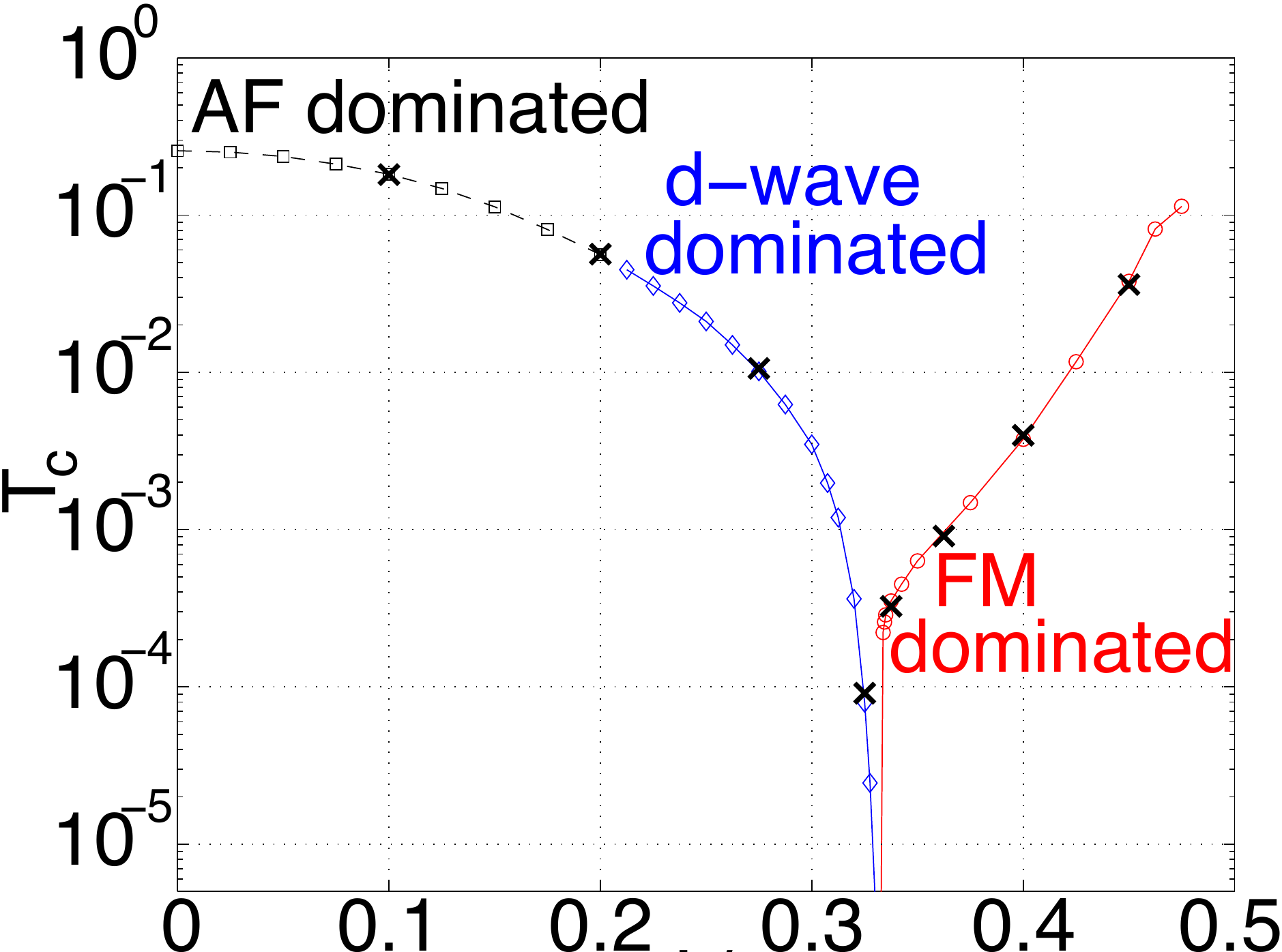}
\hfil
\includegraphics[height=0.3\textwidth]{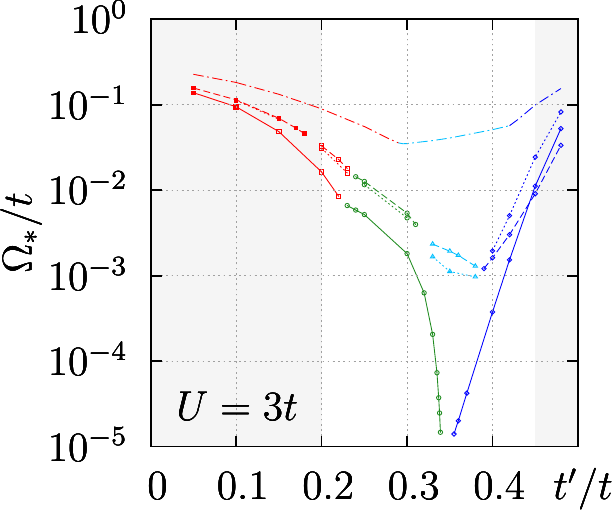}
\caption{Renormalization group results indicating a quantum critical point between a d-wave superconducting and a ferromagnetic phase in the two-dimensional Hubbard model. The parameter on the horizontal axis is $\theta = - \hopt'/\hopt$. Left: results from the temperature-flow RG in static approximation (figure taken from \cite{HSPRL}). Right: results from the $\Omega$-scheme RG with full frequency dependence taken into account (figure taken from \cite{GiSa}). The lowest curve on the right is where the frequency of both the self-energy and the interaction are taken into account, confirming the downturn of the scale where the interactions grow in the flow.}
\label{fig11}
\end{figure}

In our more recent studies \cite{GiSa}, we calculated the fully momentum- and frequency-dependent vertex functions using (\ref{eq:RGlevel2}). The results confirm the existence of a QCP. We also calculated the small-frequency behaviour of the fermionic self-energy numerically at the QCP. The result can be stated as follows.  

{\sl
\noindent
Consider a curve of $(t,t',U)$-Hubbard models parametrized by $\theta = -\frac{t'}{t}$, 
where the chemical potential $\mu(\theta)$ is chosen such that
the Fermi curve contains a Van Hove point, and $U$ is fixed. 
Then there is a point $\theta_0$, approximately given by $0.341$, where $v_\escale$ stays small down to the smallest scales $\Omega$ reachable by our numerical calculations. 
For $\theta$ close to, but smaller than $\theta_0$, the effective interaction is dominated by the $d$-wave superconducting term $D_{2,2}$. For $\theta> \theta_0$, the effective interaction is dominated by the ferromagnetic term $M_{1,1}$
At $\theta_0$, the fermionic self-energy has small-frequency behaviour
$$
\Sigma (\omega, (\pi,0)) \sim \I \; \mbox{sgn} (\omega) |\omega|^{\gamma}
$$
with $\gamma < 1$ ($\gamma \approx 0.74$).
The momentum and frequency dependence of the vertex functions is obtained as well, using the composite-field expansion introduced above, and shows similar scaling exponents. 
}

\begin{figure}[h]
\hfill \includegraphics[width=0.47\textwidth]{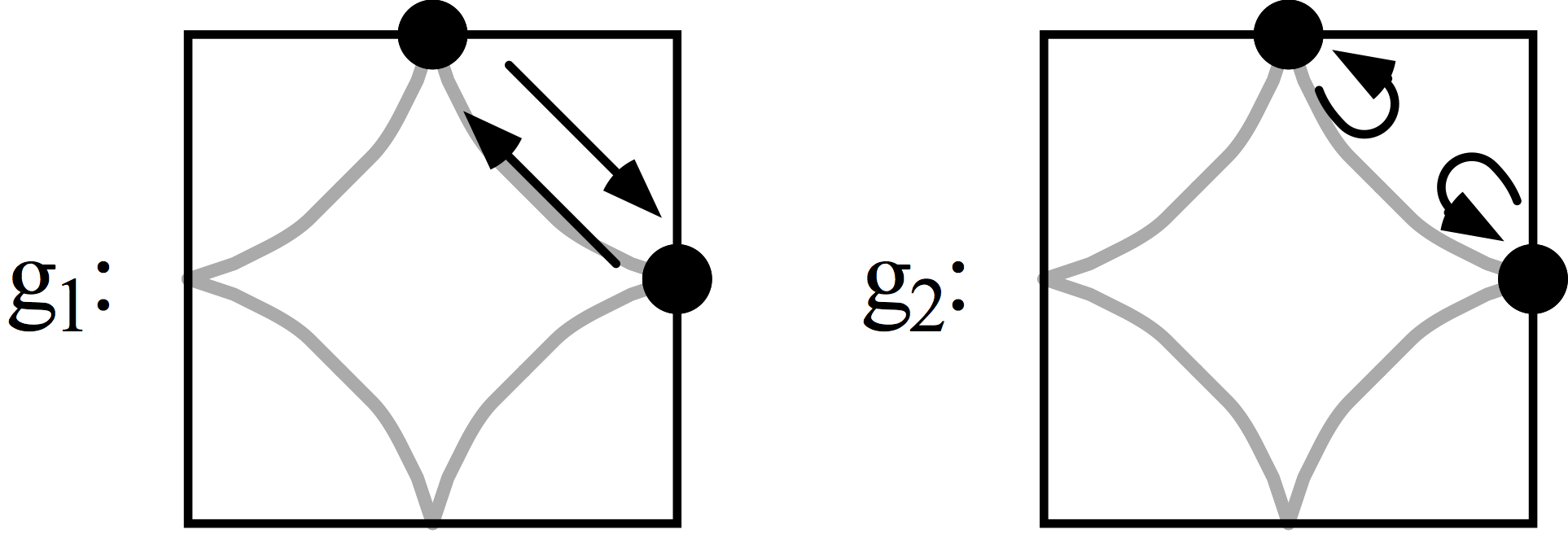}
\hfil
\includegraphics[width=0.47\textwidth]{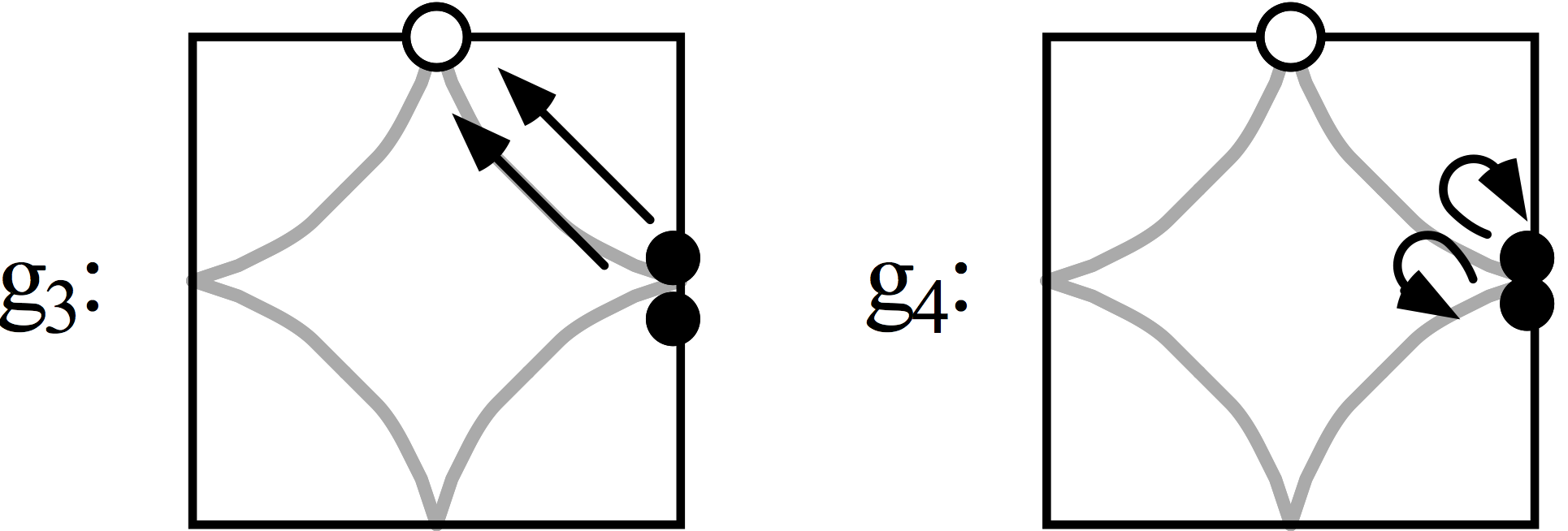}
\caption{The couplings $g_1, \ldots, g_4$ correspond to scattering processes between the points $(0,\pi)$ and $(\pi,0)$ in momentum space.
(Figure taken from \cite{FRS}.)}
\label{fig12}
\end{figure}

Although found by numerical studies, the occurrence of the QCP can be understood, at least qualitatively, by a simple analytical argument.
The special value $\theta_0$ is close to the place where two second-order contributions to the effective action cancel in second-order perturbation theory, namely the particle-hole bubble at external momentum  zero and the particle-hole bubble at external momentum $\hat \pi = (\pi, \pi)$. The significance of these two terms becomes most explicit in the so-called {\sl two-patch approximation} to the RG flow, where the coupling function is evaluated only at those momenta where singularities occur already in perturbation theory, namely $0$ and $\hat \pi$. For spin-$\frac12$ fermions, there are four couplings, corresponding to $v$ evaluated at these points (see Figure \ref{fig12}). 

\newcommand{\gone}{g_1}
\newcommand{\gtwo}{g_2}
\newcommand{\gthr}{g_3}
\newcommand{\gfou}{g_4}
\newcommand{\bubb}[2]{\dot B_{#1}^{(#2)}}
The RG equations then reduce to 
\begin{equation}
\begin{array}{lcrlcl}
\dot \gone 
&=&
- \bubb{\hat \pi}{-} 
&(\gone^2+\gtwo^2)
&+&
\bubb{0}{+} \; 2 \gfou (\gtwo-\gone) 
+
\bubb{\hat \pi}{+} \; (\gone^2+\gthr^2)
\\[1ex]
\dot \gtwo 
&=&
- \bubb{\hat \pi}{-}
& 2 \gone\gtwo
&+&
\bubb{0}{+} \; 2 \gtwo\gfou 
\mkern52mu
+
\bubb{\hat \pi}{+} \; 2 \gtwo (\gone-\gtwo) 
\\[1ex]
\dot \gthr 
&=&
- \bubb{0}{-} 
& 2 \gthr\gfou 
&+&
\bubb{\hat \pi}{+} \; 2 \gthr (2 \gone - \gtwo)
\\[1ex]
\dot \gfou 
&=&
- \bubb{0}{-} 
& (\gthr^2+\gfou^2)
&+&
\bubb{0}{+}\; (\gtwo^2 - 2 \gone^2 + \gfou^2 + 2 \gone\gtwo)
\\
\end{array}
\nonumber
\end{equation}
where 
$$
\bubb{\Sp{k}}{\pm}
=
\bubb{\Sp{k}}{\pm} (s)
=
\mp
\frac{\del}{\del s}
\int_p G_{\escale_s} (p) \; G_{\escale_s} ( (0,\Sp{k}) \pm p) \; ,
$$ 
and we have parametrized $\escale_s = \escale_0 \; \E^{-s}$, so that $\Omega \to 0$ when $s \to \infty$. The interaction term of the Hubbard model is local in position space, so its Fourier transform is independent of momentum $v^{\Omega_0} (k_1, k_2, k_3) = U$. It gives the initial condition for the $g$'s.Thus, at $s=0$, we have $\gone=\gtwo=\gthr=\gfou=U$. 

The coefficient $\bubb{0}{-}$ is positive, and it is proportional to $s$, hence gives the leading term as $s \to \infty$. Integration over $s$ gives 
\beq
B_0^{(-)} \sim s^2  = \left(\ln \frac{\Omega_s}{\Omega_0}\right)^2. 
\eeq
This behaviour corresponds to the $(\log \beta)^2$ singularity of the Cooper pair interaction. The square of the logarithm arises instead of the simple logarithm in the standard Cooper interaction because of the logarithmic divergence of the density of states at Van Hove filling. All other coefficients are smaller, but, again because of the Van Hove singularity, they remain of order $1$ for arbitrarily large $s$, thus integrating to $s = \ln \frac{\Omega_s}{\Omega_0}$, which corresponds to the $\log \beta$ singularity of magnetic interactions at Van Hove filling (which is again a consequence of the logarithmic divergence of the fermionic density of states).

Taking linear combinations of the equations for $\gthr$ and $\gfou$ gives
\begin{equation}
\begin{array}{lcrlcl}
\dot \gthr - \dot \gfou
&=&
\bubb{0}{-} 
& (\gthr- \gfou)^2 
&+&
\bubb{\hat \pi}{+} \; 2 \gthr (2 \gone - \gtwo)
\\[.5ex]
&&&
&-&
\bubb{0}{+}\; (\gtwo^2 - 2 \gone^2 + \gfou^2 + 2 \gone\gtwo)
\\[1ex]
\dot \gthr + \dot \gfou 
&=&
- \bubb{0}{-} 
& (\gthr + \gfou)^2 
&+&
\bubb{0}{+}\; (\gtwo^2 - 2 \gone^2 + \gfou^2 + 2 \gone\gtwo)
\\[.5ex]
&&&
&+&
\bubb{\hat \pi}{+} \; 2 \gthr (2 \gone - \gtwo)
\end{array}
\nonumber
\end{equation}
Initially, $\gthr-\gfou=0$, and it stays zero if no terms except the one multiplied by $\bubb{0}{-} $ were present. Then, $\gthr \to 0$ and $\gfou \to 0$ as $s \to \infty$, since $\bubb{0}{-} >0$ and $\sim s$. 
However, the additional terms influence the flow before the asymptotic region, where $\bubb{0}{-} $ dominates, is reached. For the Hubbard initial condition, we have for small $s$
\beq
\dot \gthr - \dot \gfou
\approx
\bubb{0}{-} \;
(\gthr- \gfou)^2 
+
\left( \bubb{\hat \pi}{+} - \bubb{0}{+} \right) 2 U^2
\eeq
The second term on the right hand side shifts the flow away from the line $\gthr=\gfou$. (A similar effect occurs in the equation for $\dot \gthr + \dot \gfou$.) As a consequence, both couplings can grow during the flow. Due to the quadratic nonlinearity on the right hand side of the flow equation, this leads to a blowup at a finite $s$, which signals an instability of the Fermi system. This has been analyzed in great detail, and it has been shown that flows which take symmetry breaking into account properly can be continued to zero scale and indeed lead to symmetry-broken states \cite{SHML,MSHMS}.

However, the two terms $\bubb{\hat \pi}{+}$ and $ \bubb{0}{+}$ depend on $\theta$. They are equal close to $\theta_0$, so that the additional term vanishes, and $\gthr$ and $\gfou$ remain small (see the figure, where they are calculated both in the temperature-flow scheme of \cite{HSTflo} and in the $\Omega$-scheme of \cite{HuSa}). The transition point $\theta_0$ for the full (not just 2-patch) flow differs from these crossing points, but it is essentially at the same location in both schemes. Note that for a strict momentum cutoff, $\bubb{0}{+}$ vanishes, so the effect is artificially suppressed there. 

\begin{figure}[h]
\includegraphics[width=0.9\textwidth]{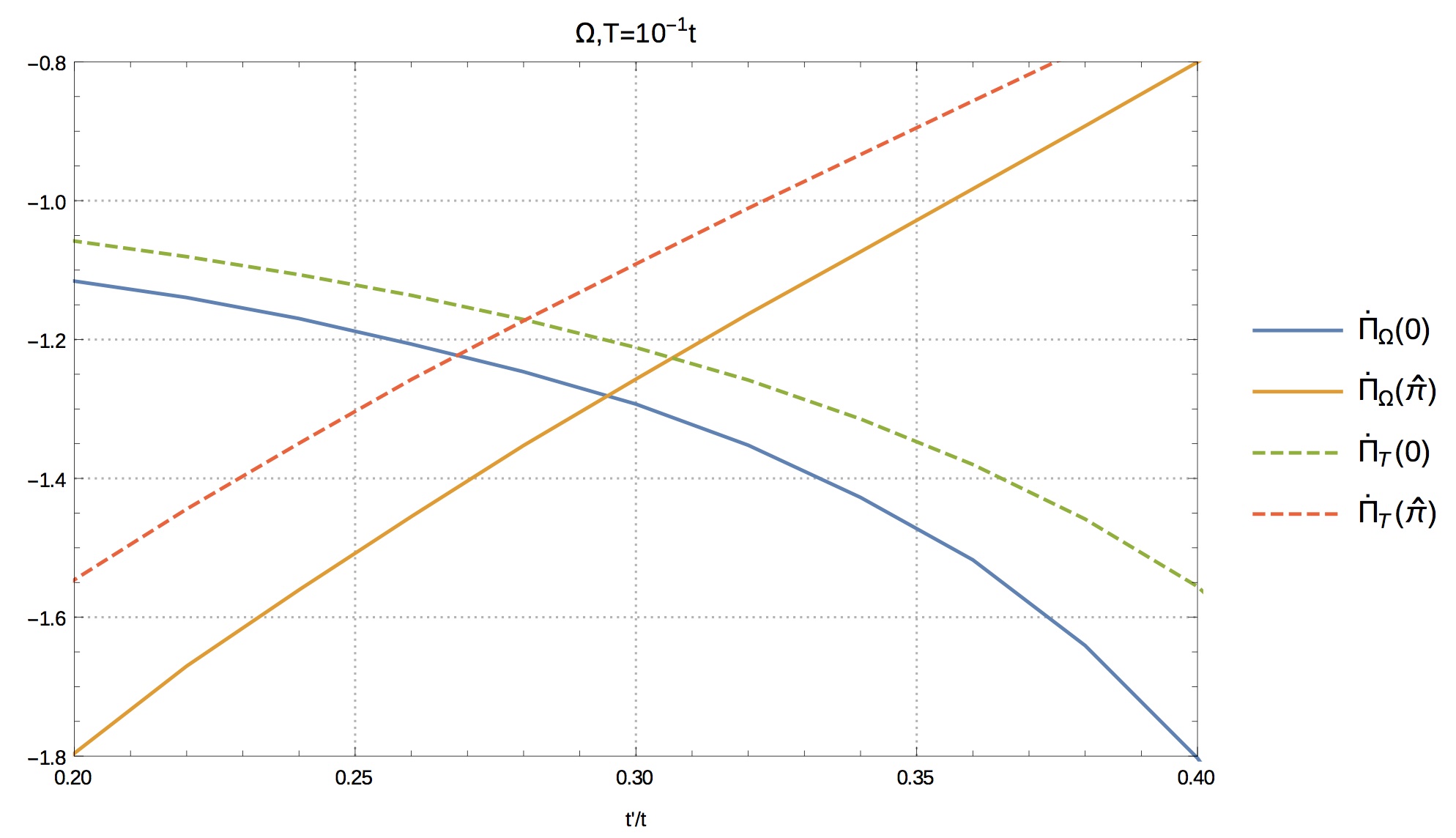}
\caption{The coefficient functions $\bubb{0}{+}$ and $\bubb{\hat\pi}{+}$ appearing in the RG equation, plotted as a function of $\theta = -\hopt'/\hopt$. Dashed lines: temperature-flow scheme, solid lines: $\Omega$-scheme. At the respective crossing points, the term driving the asymmetry between $\gthr$ and $\gfou$ cancels in the second-order diagram. In the full flow, where contributions from the entire Fermi surface and all subleading corrections are included, the transition shifts to larger $\theta \approx 0.341$, and the dependence on the schemes becomes weak (see Figure \ref{fig11}).}
\label{fig13}
\end{figure}

Thus, the Hubbard model indeed realizes the possibility of a cancellation in the flow of the two-particle interaction. Consequently, the buildup of a singularity in the frequency derivative of the self-energy remains as the leading term, and it drives the quasiparticle weight to zero at the Van Hove points. The anomalous exponent was calculated in \cite{GiSa}. The transition can be further analyzed using the effective action obtained from the RG calculations. We have checked the consistency of the anomalous exponents using an alternative calculation with Schwinger-Dyson equations. Morover, it turns out that a simple mean-field model that includes the singular fermionic self-energy can account for the downturn of critical scales \cite{KVMS-MFRG}. 

\section{Concluding Remarks}

During the last 25 years, the renormalization group has been used to achieve significant progress in understanding condensed-matter systems, in particular many-fermion systems. Rigorous mathematical control of fermionic correlation functions has been achieved using the RG together with convergent expansions. 

The conditions under which a system exhibits Fermi liquid properties, and under which it deviates from it, have in this way been clarified. For weakly coupled systems, such deviations require long-range interactions or singular Fermi surfaces.

The RG explains in a simple way how Luttinger-liquid properties arise in one spatial dimension, and why Fermi liquids can be stable in two and more spatial dimensions. It has also been instrumental in applications to models of layered materials, where it explained the occurrence of d-wave superconductivity in the repulsive Hubbard model, which is the prototypical model for the cuprate high-temperature superconductors. 

In the application part of this review, I have focused on a special, but interesting feature, namely singular self-energies due to Van Hove singularities, which drive quantum critical behaviour. While discussed specifically for Van Hove filling here, the dynamical phenomena seen here appear naturally and lead to an extension of the Van Hove region, as suggested already in \cite{Dzy96}, hence may be important in a much more general setting.

\bigskip
{\bf Acknowledgements.} I would like to thank Klaus Sibold and Wolfgang Hollik for the invitation to present a talk at the memorial symposium for Wolfhart Zimmermann. The work reported here was supported over many years by DFG Forschergruppe FOR 723: {\sl Functional Renormalization Group for Correlated Fermion Systems}.

\end{document}